\documentclass [11pt]{article}
\usepackage{graphicx}
\usepackage{amssymb} 
\usepackage{latexsym}
\usepackage{mathrsfs}
\usepackage{color}
\definecolor{red}{rgb}{1,0,0}

\makeatletter
\def\section{\@startsection {section}{1}{\z@}{-3.5ex plus -1ex minus
 -.2ex}{2.3ex plus .2ex}{\large\bf}}
\def\subsection{\@startsection{subsection}{2}{\z@}{-3.25ex plus -1ex
minus -.2ex}{1.5ex plus .2ex}{\normalsize\bf}}
\makeatother
\makeatletter

\@addtoreset{equation}{section}

\makeatother

\def\Dslash{\hspace{3pt}\raisebox{1pt}{$\slash$} \hspace{-9pt} D}
\def\Dslashn{\hspace{3pt}\raisebox{1pt}{$\slash$} \hspace{-7pt} D}
\def\bea{\begin{eqnarray}} \def\eea{\end{eqnarray}}
\def\be{\begin{equation}} \def\ee{\end{equation}} \def\nn{\nonumber}
\def\a{& \hspace{-7pt}}  \def\Z{{\bf Z}}

\begin{document}

\thispagestyle{empty}

\begin{center}
\hfill SISSA-15/2005/EP \\

\begin{center}

\vspace{1.7cm}

{\Large\bf The Electroweak Phase Transition \\[3mm]
on Orbifolds with Gauge-Higgs Unification}

\end{center}

\vspace{1.4cm}

{\bf Giuliano Panico and Marco Serone}\\

\vspace{1.2cm}

{\em ISAS-SISSA and INFN, Via Beirut 2-4, I-34013 Trieste, Italy}
\vspace{.3cm}

\end{center}

\vspace{0.8cm}

\centerline{\bf Abstract}
\vspace{2 mm}
\begin{quote}\small

The dynamics of five dimensional Wilson line phases at finite temperature is studied in the
one-loop approximation.
We show that at temperatures of order $T\sim 1/L$, where $L$
is the length of the compact space, the gauge symmetry is always restored and
the electroweak phase transition appears to be of first order.

Particular attention is devoted to the study of a recently proposed
five dimensional orbifold model (on $S^1/\Z_2$) where the Wilson line phase is
identified with the Higgs field (gauge-Higgs unification).
Interestingly enough, an estimate
of the leading higher-loop ``daisy'' (or ``ring'')
diagram contributions to the effective potential in a simple five
dimensional model,
seems to suggest that the electroweak phase transition can be
studied in perturbation theory even for Higgs masses above
the current experimental limit of 114 GeV.
The transition is still of first order for such values of the Higgs mass.
If large localized gauge kinetic terms are present, the transition
might be strong enough to give baryogenesis at the electroweak transition.

\end{quote}

\vfill

\newpage

\section{Introduction}

Theories in extra dimensions are interesting alternatives to more
standard four-dimensional (4D) extensions of the Standard Model (SM).
In particular, there has recently been a lot of interest in theories
with compact extra dimensions at the TeV scale \cite{Antoniadis}.
On the theoretical side they are, among other things, a possible framework to
solve the gauge hierarchy problem and, from a more phenomenological
point of view, they would potentially be tested at the Large Hadron Collider (LHC).

A particularly interesting scenario provided by theories in extra dimensions
is the possibility of identifying the Higgs field with the internal component
of a gauge field \cite{early}. Several models of this sort have recently been considered,
with \cite{Ghu-susy} or without \cite{Ghu-no-susy,Scrucca:2003ra}
supersymmetry, in various set-ups.
Models invoking this scenario, given the origin of the
Higgs field, are sometimes called models with gauge-Higgs unification.
Electroweak symmetry breaking occurs radiatively
in these theories, and is equivalent to a Wilson line symmetry breaking
\cite{SS,hos,Wilson-line}, since the vacuum expectation value (VEV) of the
Higgs field is proportional to the value of the Wilson line phase.
In minimal 5D non-supersymmetric single-Higgs models,
the Higgs potential is entirely radiatively generated and non-local
in the extra dimensions. The models are highly predictive and
the gauge hierarchy problem is solved.
Matter can also be introduced in this set-up, with the possibility
of naturally having a hierarchy of fermion masses \cite{Scrucca:2003ra}.
Although some problems have still to be solved before getting a completely
realistic model,\footnote{See however ref.~\cite{Agashe:2004rs}
for a potentially realistic recent proposal.}
theories with gauge-Higgs unification are definitely
an interesting framework for physics beyond the SM.

Aim of the present paper is to study in some detail the dynamics of 5D Wilson line
phases at finite temperature, focusing in particular
to the study of the Higgs potential in minimal $S^1/\Z_2$ orbifold models with gauge-Higgs unification.
By studying the one-loop effective potential, we find that the gauge symmetry broken
by the Wilson lines is always restored at temperatures of order
$T\sim 1/L$, where $L$ is the length of $S^1$, in agreement with
previous results \cite{Ho:1990xz}.
The behaviour of the effective potential at finite temperature is
different from that usually expected in 4D or higher dimensional theories.
This is best seen by looking at the dependence on the temperature of the mass
of the field representing the order parameter of the phase transition.
At high temperatures, it is known that the leading boson and fermion
contributions to that mass are proportional in 4D to $T^2$ \cite{Dolan:1973qd}, whereas
in extra dimensional theories with $n$ compact toroidal dimensions they are proportional to $(LT)^n T^2$
\cite{Dienes:1998hx}, where $L$ is the length of the extra circles, taken all equal, for simplicity.
In our model the order parameter is given by the VEV
of the Wilson line phase. Its non-local nature
is responsible for a radically different behaviour at high temperatures:
the fermion contribution is exponentially suppressed with the temperature, whereas
the boson contribution has a leading term proportional to $T/L$, plus terms
exponentially suppressed with $T$.
The mass of the Wilson line phase at zero temperature
is radiatively induced and proportional to $1/L^2$; when the latter is negative,
a symmetry breaking is generated, but then at temperatures of order $T\sim 1/L$
a phase transition occurs and the symmetry is restored. This will always happen, due to
the presence of the bosonic contribution of the 5D gauge fields.\footnote{The restoration
at high temperatures of a symmetry broken at zero temperature is not a universal feature.
It has recently been observed, for instance, that
in so called little Higgs theories the electroweak gauge symmetry remains broken even at high
temperatures, at least up to the maximum temperature which can be studied
in the effective theory \cite{Espinosa:2004pn}. In non-abelian theories, it can also happen
that the Higgs and the confining phases are connected without the appearance
of any phase transition. This is what lattice simulations predict for the SM, as
we briefly remind below.} The phase transition is analytically studied in a general framework,
in presence of bulk bosons and fermions with generic couplings to the Wilson line phase.
The transition is predicted to typically be of first order, with a strength which
decreases as the fermion couplings increase.
The appeareance of a first-order phase transition is essentially due to the presence
of a term which is cubic in the Wilson line phase. The latter arises only
from the contributions of massless 5D bosons.
The phase transition takes place at temperatures
of the same order of the compactification scale and thus well below the UV cut-off
scale $\Lambda$ of the model.

Having established the generic presence of a first-order phase transition around
$T\sim 1/L$ for Wilson line based models, we study some properties of
the electroweak transition occurring in 5D
models of gauge-Higgs unification.
For definiteness, we focus to the particular model (and its generalizations)
proposed in ref.~\cite{Scrucca:2003ra}.

Before summarizing our results, it might be useful
to review very briefly the nature of the electroweak phase transition --- and how it can reliably be
studied --- in the SM \cite{Kirzhnits:1972ut}.
The latter properties are known to essentially depend on the value of the Higgs quartic coupling
$\lambda$, or equivalently on the Higgs mass $M_H$.
For $M_{H}< M_W$,
the transition is always of the first order, with a strength
that is inversely proportional to the Higgs mass \cite{Carrington:1991hz,Arnold:1992rz}.
In this regime, the SM phase transition can
reliably be computed in perturbation theory, although the resummation of an infinite class of
certain IR divergent higher loop diagrams (so called ``daisy'' or ``ring'' diagrams) has to be performed.
Also in the SM case, the presence of a first-order phase transition can be traced back to the presence
of a term in the potential which is cubic in the Higgs field.
Since the effective potential is generally a gauge-dependent quantity, care has to
be taken in the gauge-fixing procedure \cite{Laine:1994bf} or a gauge-invariant formalism
has to be advocated \cite{Hebecker:1994}.
Around the critical temperature, perturbation theory is less and less reliable
as $M_{H}$ approaches $M_W$.
For $M_{H}\gtrsim M_W$, perturbation theory breaks down
and one has to consider non-perturbative methods, such as lattice computations.
The latter seem to indicate that in this regime of Higgs masses, the SM does not actually
have a phase transition but rather a crossover, since the Higgs and the confining phase
are continuously  connected \cite{Kajantie:1996mn}.

One of the main motivations to study the detailed nature of the electroweak
phase transition in the SM has been the realization that, under certain assumptions
(namely a sufficiently strong CP violation and first order phase transition), it could
satisfy the necessary requirements for baryogenesis \cite{Baryo}.
Given the actual bounds on the value of the Higgs mass and amount of CP violation in the SM,
it has definitively established that the SM phase transition is unable to give a successful
baryogenesis.\footnote{Proposals to improve the SM situation introducing effective dimension
six $(H^\dagger H)^3/M^2$ operators \cite{Dim6} or strongly coupled fermions \cite{Carena:2004ha}
have recently appeared.}

Let us now come back to analyze our results. In the model discussed in ref.~\cite{Scrucca:2003ra},
the Higgs mass is predicted and its value is typically $M_H< M_W/2$.
In this regime, it is reasonable to compare our results with those obtained in the SM
at one-loop level, for the same value of Higgs and top mass.\footnote{The top quark mass
in the model of ref.~\cite{Scrucca:2003ra} is typically
lower than the actual experimental one, $M_{top} \simeq 174$ GeV.}
We find that the electroweak phase transition is of first order,
with a strength that is inversely proportional to the Higgs mass,
similarly to what happens in the SM (see figures \ref{figRadiusBFerm1}-\ref{figEffPotBFerm1}).

When 5D bulk fermions in very large representations of the gauge group are inserted,
so that realistic values of the Higgs mass can be obtained, the first order phase transition
becomes weaker and weaker
(see figures \ref{figRadiusRank8}-\ref{figEffPotRank8}).
On the other hand, one can get realistic values of the Higgs (and top) mass also by
introducing large localized gauge kinetic terms, as discussed in ref.~\cite{Scrucca:2003ra}.
In this case the strength of the phase transition
is considerably increased (see figures \ref{figRadiusKin1}-\ref{figEffPotKin1}) and could be strong enough
to produce a reasonable baryogenesis.
A definite result can however be established only by a more careful
study of the phase transition, which includes the study of sphalerons
and of the possible sources of CP violation in our 5D set-up.

As we have already mentioned, all our results, for any value of $M_{H}< M_W$ or
$M_{H}\geq M_W$, are based on perturbative studies, obtained by computing the
one-loop Higgs effective potential. In light of the importance of the correct gauge-fixing
procedure, of the higher-loop daisy contributions and of the breakdown of perturbation theory for
$M_{H}\gtrsim M_W$ in the SM, it is natural to ask if and to what extent one can trust
our perturbative one-loop results. To this aim, we show that our 5D one-loop Higgs effective
potential is considerably stable under radiative corrections. First of all,
we establish that the 5D one-loop Higgs effective potential
is gauge independent, by studying a class of background field gauges. This is expected,
since the effective action typically fails to be gauge invariant away from
its minima, but at tree-level in our model any VEV of the Higgs field is a good minimum,
being absent any potential at this order.

We give support to the idea that the phase transition can safely be studied in perturbation theory
in Wilson line based models by
computing the leading IR divergent daisy contributions to the Wilson line effective potential
in a simple 5D model compactified on $S^1$. By a comparison between the na\"\i ve one-loop
thermal mass with the one-loop improved one --- obtained by resumming an infinite class
of daisy diagrams --- we are able to establish that
the leading daisy diagrams give a small correction up to temperatures of order $T\sim 1/L$
(see figure 19 (a)) and start to give a
relevant contribution only at temperatures of order $T\sim 10/L$ or higher.
We take these results as a strong evidence that perturbation theory in 5D models
with gauge-Higgs unification is valid also around the critical temperature.
The same comparison of one-loop na\"\i ve and improved thermal masses
is performed in the 4D theory obtained by a trivial dimensional reduction of the
5D model, where only the Kaluza-Klein (KK) zero modes of the fields are retained.
This truncation spoils the higher dimensional symmetries responsible
for the finiteness of the Wilson line effective potential and indeed, as expected,
daisy diagram contributions are larger (see figure 19 (b)).

The paper is organized as follows. In section 2 we show the main features of
one-loop effective potentials of Wilson line phases,
namely their UV finiteness, gauge independence, high temperature behaviour
and establish the generic presence of a first-order phase transition.
In section 3 we study the electroweak phase transition in the model of ref.~\cite{Scrucca:2003ra}, which we
briefly review in subsection 3.1. In section 4, the leading higher loop daisy
diagrams in an $U(1)$ model coupled with scalars and compactified on $S^1$ (and its
corresponding model trivially reduced in 4D) are computed. Our conclusions are
reported in section 5.
Finally, we collect in an appendix all the formulae regarding the contribution to the
Wilson line effective potential of bulk fields, with or without possible localized
couplings.

\section{The Wilson line dynamics at finite temperature}\label{secFiniteTempPot}

The form of the one-loop effective potential at $T=0$ for Wilson line
phases on $S^1$ or $S^1/\Z_2$ orbifold models
is well-known \cite{hos,Kubo:2001zc}.
The non-local nature of Wilson lines imply that
their potential is completely UV finite, modulo irrelevant constant
vacuum energy terms. This is the most important property
of these potentials, which makes the idea of identifying
the Higgs field with a Wilson line phase such an interesting
proposal to solve the gauge hierarchy problem.
We consider below only the effective potential for a single Wilson
line phase. Effective potentials for multiple Wilson line phases, however,
do not present any problem and are easily obtained by a generalization
of that of a single Wilson line phase (see {\em e.g.} refs.~\cite{hos,Haba:2002py,Haba2}).

In this class of models, gauge invariance
severely constrains the couplings of the gauge field component $A_5$ with other fields.
In particular, any bulk field couples to $A_5$
only through the minimal gauge coupling.
In a suitable basis for the fields, each mode $(i)$ couples diagonally
with $A_5$, giving rise to a simple shift of their KK mass
(see {\em e.g.} appendix A of ref.~\cite{Scrucca:2003ra} for details; all our conventions
and notations are as in ref.~\cite{Scrucca:2003ra}):
$m^{(i)} = (n + q^{(i)}\alpha)/R$, where $n$ is the KK level of the
state, $q^{(i)}$ is the charge eigenvalue
of the mode $(i)$ and $\alpha$ ($0\leq \alpha \leq 1$) is the VEV
of the Wilson line phase. It is related to the zero mode $H$ of $A_5$
by the relation $H=2i\alpha/(g_4 R)$,\footnote{The precise coefficient relating $\alpha$
and $H$ is model dependent. For definiteness, we have taken here and in the following
the one appearing in ref.\cite{Scrucca:2003ra}.}
where $g_4$ is the 4D gauge coupling constant, related to the
dimensionful 5D gauge coupling constant as follows: $g_4=g_5/\sqrt{L}$, with
$L=2\pi R$.

The effective potential at finite temperature, in the imaginary
time formalism, is obtained from that at zero temperature by simply
compactifying the Euclidean time direction on a circle of radius
$1/(2\pi T)$, where $T$ is the temperature.
Fields satisfying Bose/Fermi statistic must be taken to
satisfy periodic/anti-periodic boundary conditions along the
Euclidean time direction. The bosonic and fermionic contributions to
the one-loop Wilson line effective potential
at finite temperature are then easily evaluated. For each bosonic and fermionic
degree of freedom with charge $q$ and bulk mass term $M$, one has
\be
V(T,q\alpha) = \frac{(-)^{2\eta}T}{2}\sum_{k,m=-\infty}^{+\infty} \int\!\frac{d^3p}{(2\pi)^3}
\log \bigg\{ p^2 + \bigg[2\pi (m+\eta) T\bigg]^2 + M_k^2(q \alpha)\bigg\} \,,
\label{Pot0}
\ee
where $M_k^2(q \alpha)=M^2 + (k+q\alpha)^2/R^2$ are the masses of the states at KK level $k$,
including a possible bulk mass term $M$,
$m$ are the Matsubara frequencies and $\eta=0$ or 1/2 for bosons and fermions, respectively.
It is important to recall that eq.(\ref{Pot0}) is valid not only for compactifications on $S^1$
but also on orbifolds, such as $S^1/\Z_2$. In the latter case, the effect of the parity
projection in the diagonal basis for the fields amounts simply to
reducing the number of physical degrees of freedom
with a given charge $q$ with respect to the Wilson line phase.
We collect in the appendix several equivalent ways in which one can compute the effective
potential (\ref{Pot0}). Notice that eq.(\ref{Pot0}) is valid also for gauge fields $A_M$ $(M=0,1,2,3,5)$;
in this case one has to take into
account a multiplicity factor equal to 3, the number of transverse polarization of a 5D gauge
field. Moreover, the gauge contribution is gauge independent.
We will discuss in more detail this important aspect in next subsection.

\subsection{Gauge independence of the effective potential}

The effective potential $V(H_0)$ in standard spontaneously broken gauge theories,
such as the SM, where $H_0$ is the VEV of the Higgs field,
is generally gauge-dependent \cite{Dolan:1973qd}. This is easily seen at one-loop level,
where the effective potential is obtained by evaluating a trace.
In this case one can see, working for instance in an $R_\xi$-gauge,
that the contributions of the would-be Goldstone bosons, ghosts and longitudinal gauge bosons,
outside the minimum for the Higgs field, do not cancel anymore and leave a non-trivial
$\xi$-dependent, and thus gauge dependent, term.

In gauge-Higgs unification models, where the Higgs field is identified with the KK zero mode
of the gauge field $A_5$,
such problem does not occur, at least at one-loop level. This is expected since at tree level
$A_5$ is a modulus, namely any constant value of this field satisfies the equations of motion.
In order to show more explicitly the gauge independence of the one-loop effective potential,
we define in the following a class of gauge-fixing Lagrangians ${\cal L}_\xi$ that are
a sort of generalization of the 4D t'Hooft background gauges:
\be
{\cal L}_\xi = -\frac 1{\xi} {\rm Tr}\, \big( \bar D_\mu A^\mu + \xi \bar D_5 A^5 \big)^2\,,
\label{L-gf}
\ee
where $\mu=0,1,2,3$ runs over the 4D directions and
$\bar D_M = \partial_M + i g_5 [\bar A_M,.]$ is the covariant derivative in terms of
the classical field configuration $\bar A_M =\delta_{M5}\alpha/(g_5 R) \hat t$, $\hat t$ being
the direction in group space where the VEV is aligned.
For $\xi=1$, the
gauge-fixing term (\ref{L-gf}) reduces to the usual 5D background field gauge commonly used
in the literature, to derive the one-loop effective potential of Wilson line phases \cite{hos};
for $\xi=0$, we instead get the 4D Landau gauge $\partial_\mu A^\mu =0$, whereas
for $\xi\rightarrow \infty$, eq.(\ref{L-gf}) gives $\bar D_5 A^5=0$;
the latter condition is precisely the unitary gauge in which all would-be Goldstone
bosons associated to the non-linearly realized gauge symmetries are decoupled, as we will see
more explicitly in the following.
{}From a 4D point of view, eq.(\ref{L-gf}) is precisely the gauge-fixing term
defining an $R_\xi$-gauge for the infinite gauge symmetry groups associated to all
KK levels of the 5D gauge fields $A_\mu(x,y)$.
The full 5D Lagrangian for the gauge fields is then
\be
{\cal L}=-\frac 12 {\rm Tr} \,\big(F_{MN} F^{MN}\big)+ {\cal L}_\xi + {\cal L}_{gh}\,,
\ee
where ${\cal L}_{gh}$ is the ghost Lagrangian associated to the gauge-fixing (\ref{L-gf}):
\be
{\cal L}_{gh} = - {\rm Tr}\, \big[\omega^*  ( \bar D_\mu D^\mu + \xi \bar D_5 D^5) \omega\big]\,.
\ee
It is not difficult to derive the quadratic Lagrangian, in momentum space,
for the gauge, scalar and ghost fields $A_\mu$, $A_5$ and $\omega$.
As discussed before, the only non-trivial task is to derive the linear
combination of fields that couples diagonally with $A_5$, in which base
the effect of the background is simply to shift the KK level of the mode.
For definiteness (and because this will be the case we analyze mostly),
we consider an $SU(3)$  gauge theory on the orbifold $S^1/\Z_2$, where
$SU(3)$ is broken by the orbifold parity to $SU(2)_L\times U(1)_Y$ and thus to
$U(1)_{EM}$ by the Wilson line phase $\alpha$.
The quadratic Lagrangian of this model reads, in momentum space,
\bea
{\cal L}_{quad.} = &&  \sum_{i=1,2}\sum_{n=-\infty}^{+\infty} \bigg[A_{\mu,n}^{(i)^\dagger}
\Delta_{\mu\nu,n}^{(i)} A_{\nu,n}^{(i)} + A_{5,n}^{(i)^\dagger}
\Delta_{55,n}^{(i)} A_{5,n}^{(i)}+
\omega_{n}^{(i)\dagger} \Delta_{55,n}^{(i)} \omega_{n}^{(i)} \bigg] + \nn \\
&& \sum_{i=3,4}\sum_{n=0}^{+\infty} \bigg[A_{\mu,n}^{(i)}
\Delta_{\mu\nu,n}^{(0)} A_{\nu,n}^{(i)} + A_{5,n}^{(i)}
\Delta_{55,n}^{(0)} A_{5,n}^{(i)}+
\omega_{n}^{(i)*}
\Delta_{55,n}^{(0)} \omega_{n}^{(i)}
\bigg]\,.
\label{Lquad-espl}
\eea
In eq.(\ref{Lquad-espl}), the fields $(A_\mu^{(1,2)},A_5^{(1,2)},\omega^{(1,2)})$
represent the linear combination of fields that couples diagonally to the Higgs field,
with charge $q=1,2$. The fields $(A_\mu^{(3,4)},A_5^{(3,4)},\omega^{(3,4)})$, instead,
do not couple to $A_5$. The explicit form of these linear combinations of fields,
in terms of the usual $SU(3)$ Gell-Mann field decompositions, can be found
in Appendix A.3 of ref.~\cite{Scrucca:2003ra}.  Notice that due to the different $\Z_2$ parities
of $A_\mu$ and $A_5$, we have $A_{5,0}^{(3)}=A_{\mu,0}^{(4)}=\omega_0^{(4)}=0$.
The inverse propagators are
\bea
\Delta_{\mu\nu,n}^{(q)} = && \eta_{\mu\nu} \bigg[ p^2
+ \frac{(n+q\alpha)^2}{R^2} \bigg]
+\frac{1-\xi}\xi p_\mu p_\nu\,, \ \ \ \ \ \ \ q=0,1,2, \nn \\
\Delta_{55,n}^{(q)} = && p^2 + \xi
\frac{(n+q\alpha)^2}{R^2}\,, \hspace{3.8cm} q=0,1,2,
\label{Propagators}
\eea
where $p^2=p_\mu p^\mu$.
It is clear from eq.(\ref{Propagators}) that when $\xi\rightarrow \infty$,
all scalar fields $A_{5,n}$ decouple from the theory,
with the only exception of $A_{5,0}^{(4)}$, namely the Higgs field.
It is now easy to establish the independence on the parameter $\xi$ of the
one-loop effective potential. Indeed, for each KK mode level $n$ and charge $q$,
the gauge, scalar and ghost contributions give a total factor
\be
\frac{\Delta_{55,n}^{(q)}}
{\bigg[ \Delta_{55,n}^{(q)}\,{\rm det}\, \Delta_{\mu\nu,n}^{(q)} \,\bigg]^{1/2}}
= \sqrt{\xi} \bigg(p^2+\frac{(n+q\alpha)^2}{R^2}\bigg)^{-\frac{3}2}\,,
\label{factor}
\ee
where the determinant in eq.(\ref{factor}) is meant
to be taken only on the $4\times 4$ polarization matrix of the gauge fields.
Aside from the irrelevant $\sqrt{\xi}$ factor, which is reabsorbed in the functional
measure, all the non-trivial $\xi$-dependence in eqs.(\ref{Propagators})
cancels out in eq.(\ref{factor}), leaving an effective contribution of three
scalar degrees of freedom with twist $q$. As stated before, this equals
the total contribution of a 5D gauge field $A_M$ with the same twist.

The gauge independence of the potential
is trivially extended at finite temperature. Indeed, in the imaginary time formalism,
eq.(\ref{factor}) still holds, the only modification being given by the Matsubara
frequency modes, that are all integer-valued for gauge, scalar and ghost fields.

\subsection{High temperature behaviour}

The one-loop effective potential for Wilson line phases at finite temperature has been
computed in  ref.~\cite{Ho:1990xz} (see also  ref.~\cite{Shiraishi:1986wu}), where it has been
shown that at temperatures of order $T\sim 1/L$ the symmetry is restored.
Its high temperature behaviour is peculiar, as can be seen by studying
the bosonic and fermionic contribution to the Higgs thermal mass at $\alpha=0$:
\be
M_H^2(T,\alpha=0)= \bigg(\frac{g_4 R}2\bigg)^2 \,
\frac{\partial^2 V}{\partial \alpha^2}\bigg|_{\alpha=0} \;.
\label{MH-0}
\ee
For $T \gg 1/L$, the mass (\ref{MH-0}) can easily be extracted from
eq.(\ref{Pot2}), using the known asymptotic behaviour of Bessel functions for large values of
their argument.
For massless bosonic and fermionic bulk fields ($M=0$ in eq.(\ref{Pot2})) with twist $q$,
one finds respectively\footnote{Recall that theories in extra dimensions
are non-renormalizable
and thus are effective theories valid up to an UV cut-off scale $\Lambda\gg 1/R$. For this reason,
one cannot consider temperatures larger than (or of the same order of) $\Lambda$,
because at these scales the theory becomes strongly coupled.\label{footCutOff}}
\bea
M_H^2(T,0)= && \frac{g_4^2}{16\pi^2} \left(\frac 23 \pi^2 q^2 n_B \right) \frac{T}{L}
\bigg\{ 1 + O\bigg[(LT)^{\frac 32} e^{-2 \pi LT}\bigg] \bigg\}
\,,
\label{thermal-mass-bos} \\
M_H^2(T,0)= && - \frac{g_4^2}{16\pi^2}  \left(2 \sqrt{2} \pi^2 q^2 n_F \right)
\frac{T}{L}(LT)^{\frac 32} e^{-\pi LT}
\bigg[ 1 + O\left(\frac{1}{LT}\right)\bigg]\,,
\label{thermal-mass-fer}
\eea
where $n_B$ and $n_F$ denote the effective number of bosonic and fermionic degrees of freedom
of the field.
The exponential terms appearing in eqs.(\ref{thermal-mass-bos}), (\ref{thermal-mass-fer})
and the linear temperature dependence of the boson contribution are
a characteristic feature of the non-local nature of the Higgs potential.
They are different from the typical contributions expected for models in extra dimensions
where one gets a thermal mass of order $L T^3$ \cite{Dienes:1998hx}, or for
the 4D Standard Model, in which the Higgs thermal mass,
at high temperatures, goes like $T^2$.

The leading high temperature dependence of the one-loop thermal mass can
be understood by studying the UV behaviour of the one-loop mass correction at zero temperature.
Since this is in general quadratically divergent, power counting
implies that its leading finite temperature contribution will be
of order $T^2$. This counting can also be used in models in extra dimensions.
Since at a given $T$, the effective number of KK modes
contributing to the thermal ensemble is roughly given by $LT$,
one would generally get a leading high temperature
contribution to the thermal mass of order $(LT) T^2=LT^3$,
in agreement with  ref.~\cite{Dienes:1998hx}.
For Wilson line potentials, however, the analysis is different, since they
are finite at zero temperature, with a
one-loop mass saturated by the compactification scale $1/L^2$, times $g_4^2/(16\pi^2)$,
a 4D loop factor.
The non-local nature of the potential is such that all Matsubara mode contributions, but
the zero mode one, are exponentially suppressed at high temperatures. The only relevant
contribution arises then from bosons, which admit a zero Matsubara mode.
Consequently, the thermal mass has a linear dependence on $T$,
reproducing the leading term in eq.(\ref{thermal-mass-bos}).
Since the effective number of KK modes contributing in the thermal ensemble
is still $LT$, eq.(\ref{thermal-mass-bos}) can alternatively be understood as the sum of the
one-loop, zero temperature, mass corrections of the $(LT)$ KK modes:
$(LT)/L^2 = T/L$.
The na\"\i ve power counting argument, applied instead to a truncated sum of KK modes,
would predict the wrong $LT^3$ dependence. In fact, a na\"\i ve truncation of the
KK sum would spoil the shift symmetry $\alpha\rightarrow \alpha + 1$ of the Wilson line
phase, resulting also in a fake divergent
zero temperature effective potential.

The high T contribution to the Higgs thermal mass of massive bulk fields, with mass $M$,
is further exponentially suppressed. In particular, for $M\gg 1/L$,
the linear dependence on $T$ appearing in the first
term in eq.(\ref{thermal-mass-bos}) is suppressed by a factor $e^{-ML}$. The same
is true for the whole effective potential, as is clear from eq.(\ref{Pot2}).
The effective potential is thus exponentially insensitive to UV physics and completely
determined by the light 5D degrees of freedom.

Given the high $T$ behaviour of the bosonic and fermionic contributions to the
Higgs mass in eqs.(\ref{thermal-mass-bos}) and (\ref{thermal-mass-fer}),
we can establish on quite general grounds that if a gauge symmetry breaking occurs
at $T=0$, namely $M^2(T=0,\alpha=0)<0$, it will be restored at a sufficiently high
temperature when $M^2(T,\alpha=0)\geq 0$. Since Wilson line
effective potentials are gauge invariant,
$V(-\alpha)=V(\alpha)$, and thus $\alpha=0$ is always an extremum of the potential.
Consequently, at sufficiently high $T$, $\alpha=0$ turns to
a minimum of the potential, at least locally.

\subsection{Properties of the phase transition}

An analytical study of the nature of the phase transition occurring
in Wilson line based models is not totally straightforward, due to the
non-local nature of the order parameter. The essential features of the transition, however,
can quantitatively be studied by properly approximating the exact
one-loop formulae for the potential appearing in the appendix.

As we will show, the phase transition is typically of first order and
is related to the presence of a cubic term in the bosonic contribution
of the effective potential. This is very similar to what happens, for instance,
in the SM, where a cubic term in the Higgs field appearing in the effective
potential at finite temperature is responsible for the generation of a first-order
phase transition, for sufficiently low values of the Higgs mass.
The cubic term in the Wilson line phase $\alpha$ appears from a high temperature
expansion of the bosonic contribution in, say, eq.(\ref{Pot2}).
For massless fields with $M=0$, charge $q_B$ and for $LT > 1$, we can safely drop all Matsubara modes
but the zero mode in the sum over $m$ appearing in eq.(\ref{Pot2}). For each degree of
freedom, the resulting potential is
\be
V_B(T,\alpha) \simeq -\frac{T}{\pi^2 L^3}
\sum_{\tilde k=1}^\infty
\frac{\cos(2\pi \tilde k q_B \alpha)}{\tilde k^{4}}\,.
\label{Pot2-App}
\ee
Interestingly enough, the potential (\ref{Pot2-App}) admits a very simple
polynomial expression in $\alpha$, which can be found by carefully Taylor
expanding it in powers of $\alpha$.
Since a na\"\i ve expansion in $\alpha$ would lead to the appearance
of an infinite number of ill-defined sums, a regularization has to be taken
to give a meaning to the resulting expression. We will use in the following
the $\zeta$-function regularization, according to which
$\zeta(-2n)=\sum_{k=1}^\infty k^{2n}=0$ for all positive integers $n$.
As a consequence, in the expansion of the cosine function in eq.(\ref{Pot2-App}),
all terms but the first three, constant, quadratic and quartic in $\alpha$, vanish.
All the odd derivatives in the expansion vanish since $\zeta(-2n+1)$ is finite for $n>0$,
but care has to be taken for the third derivative.
The latter in fact does not vanish, due to the relation
\be
\lim_{x\rightarrow 0^+}\sum_{n=1}^{\infty} \frac{\sin(n x)}{n} = \frac{\pi}{2}\,.
\label{cubic}
\ee
This is at the origin of the $\alpha^3$ term in the expansion of the potential
(\ref{Pot2-App}). It only arises for massless 5D bosonic fields, since
for massive fields the sum over $n$ is absolutely convergent, resulting
in a trivially vanishing third derivative of the potential.

Putting all terms together and
recalling that $\zeta(0)=-1/2$, one gets
\be
V_P(T,\alpha) = -\frac{\pi^2 T}{3 L^3} \bigg[ \frac{1}{30} - (q_B \alpha)^2
+2 (q_B \alpha)^3 - (q_B \alpha)^4 \bigg]\,.
\label{Pot-An}
\ee
The expressions (\ref{Pot2-App}) and (\ref{Pot-An}) are identical for $0\leq \alpha\leq 1/q_B$.
Since $V_P(T,q_B\alpha=0)=V_P(T,q_B\alpha=1)$, by periodically
extending $V_P$ to all values of $\alpha$, one gets a complete identity between the two expressions.
Indeed, it is straightforward to check that the Fourier series of the polynomial
(\ref{Pot-An}) agrees with eq.(\ref{Pot2-App}).
Notice that the $\alpha^3$ term is a non-analytic term, since it corresponds to a
$(H^\dagger H)^{3/2}$ operator, when written in a manifestly gauge-invariant form.

Let us now study how the phase transition schematically occurs in a model with one massless 5D
gauge boson and fermion, with charge respectively $q_B$ and $q_F$ with respect to $\alpha$.
We take $q_F>q_B$, because in this case the analysis is further simplified, as we will see.
There is no need to specify, in this simple set-up, which is the underlying theory
which gives rise to these fields and to the Wilson line. As a consequence,
the analysis is general and it can be applied for any gauge group
and compact space ($S^1$ or $S^1/\Z_2$).

At zero temperature, the effective potential is given by the first row in eq.(\ref{Pot3})
with $M=0$. By summing the fermionic and bosonic contributions, we get
\be
V(0,\alpha) = \frac{3}{4 \pi^2 L^4}
\sum_{\tilde k=1}^\infty
\frac{\big[8\cos(2\pi \tilde k q_F \alpha) -3\cos(2\pi \tilde k q_B\alpha)\big]}{\tilde k^{5}}\,.
\label{Pot2-0T}
\ee
The potential is unstable at $\alpha=0$ and develops a minimum
at $\alpha \simeq 1/(2q_F)$, where the ``Higgs mass'' is approximately given by
\be
M_H^2 \simeq \frac{g_4^2}{16\pi^2} \frac{1}{L^2} 24 q_F^2 \zeta(3)\,,
\label{Higgs}
\ee
where, as rough approximation, we have neglected the bosonic contribution in eq.(\ref{Higgs}).
At finite temperature, as we have already discussed, the fermion contribution
is less and less relevant, resulting in a phase transition for $T\sim 1/L$.
The transition is more efficiently studied by starting from temperatures
slightly above the critical one, when $TL>1$. In this regime, it is a good
approximation to use the high temperature expansion for the potential.
The bosonic contribution has been already computed and gives rise to
the polynomial potential (\ref{Pot-An}) multiplied by a multiplicity factor of 3.
The fermionic contribution, unfortunately, does not admit a simple finite polynomial expansion
which accurately approximates the exact one-loop result, like the bosonic case.
As such, a rough approximation will be used to put it in a simple polynomial form.
By expanding the modified Bessel functions
and retaining only the modes $m=0,-1$ and $\tilde k=1$ in eq.(\ref{Pot2}), we roughly get
\be
V_F(T,\alpha)\simeq \frac{4\sqrt{2}}{L^4} (LT)^{5/2} e^{-\pi LT} \cos(2\pi q_F\alpha)\,.
\label{Pot-AFer}
\ee
No $\alpha^3$ term arises from the fermion contribution.
If we expand the cosine in eq.(\ref{Pot-AFer}) and retain only terms up to
$\alpha^4$ we see that the effective potential for $LT>1$, resulting by summing $V_B$ and $V_F$,
has the schematic form\footnote{As we said, this is generally a rough approximation but it is enough
for our analytical estimates. All our results in the following section are
instead based on numerical analysis which take into account the exact
form of the one-loop boson and fermion contribution to the potential, as reported in the appendix.}
\be
\frac{L^4}{\pi^2} V(T,q\alpha) \simeq a(x) \,\alpha^2 - b(x) \, \alpha^3 + c(x) \,\alpha^4\,,
\label{Pot-gen}
\ee
where $x=LT$ and
\bea
a(x) & = & q_B^2 x - 8 q_F^2 \sqrt{2} x^{5/2}e^{-\pi x}\,, \nn \\
b(x) & = & 2 x q_B^3 \,, \label{a-b-c} \\
c(x) & = & q_B^4 x + \frac83 q_F^4 \sqrt{2}\pi^2 x^{5/2}e^{-\pi x} \,. \nn
\eea
Eq.(\ref{Pot-gen}) is valid for $0\leq \alpha \leq 1/(2q_F)$, which is the
relevant range in $\alpha$ for the study of the phase transition.
The analysis of the latter for potentials like eq.(\ref{Pot-gen}) is
standard (see {\em e.g.} ref.~\cite{Anderson:1991zb}). For $x>>1$, the potential admits only
a minimum at $\alpha=0$. As the temperature decreases, an inflection point appears at $T=T_1$,
below which
a non-trivial minimum and maximum appear.
The critical temperature $T_C$ is defined as the temperature when $V(T_C,q\alpha_{min}(T_C))=0$,
namely when the non-trivial minimum is as deep as the minimum at zero.
This is given by the largest root of the equation $b^2(x_C)- 4 a(x_C) c(x_C)=0$ which,
for $q_F>q_B$, is nearly equivalent to the vanishing of the term $a(x)$ in eq.(\ref{Pot-gen}).
The critical temperature has only a logarithmic dependence on the charges $q_B$ and $q_F$
and is of order $1/L$, as mentioned.
At $T=T_C$, we get
\be
\alpha_{min}(x_C) = \frac{b(x_C)}{2 c(x_C)}\simeq \frac{6 q_B}{\pi^2 q_F^2}\,.
\label{alfa-min}
\ee

Below the critical temperature $T_C$, $\alpha_{min}(T)$ becomes the new global minimum of the potential.
The value $|H(T_C)|/T_C = 2\alpha_{min}(T_C) /(g_4 R T_C)$
is one of the relevant parameters to study, if one wants to get baryogenesis
at the electroweak phase transition (see {\em e.g.} ref.~\cite{Quiros:1999jp}).
One necessary (but not sufficient) requirement
is that $|H(T_C)|/T_C> 1$, otherwise sphalerons at $T<T_C$ would wash out any previously
generated baryon asymmetry \cite{Baryo}.\footnote{The bound $|H(T_C)|/T_C> 1$
is derived by an analysis of sphaleron dynamics
in the SM. We have not repeated this analysis for Wilson lines in 5D and thus deviations
from the bound $H(T_C)/T_C> 1$ could be present.\label{footBaryonAsymmetry}}
It is also a good parameter to measure the strength
of the first-order phase transition.
Since the Higgs mass (\ref{Higgs}) is proportional to $q_F$,
whereas $\alpha_{min}(x_C)\sim q_B/q_F^2$ and $T_C\sim 1/L$, we conclude that
\be
\frac{|H(T_C)|}{T_C} \sim \frac{q_B}{q_F^2}\,.
\label{H-T}
\ee
The strength of the first-order phase transition is inversely proportional to $q_F^2$
and hence to the value of the Higgs mass.

In presence of several massless bosonic and fermionic fields, with charges $q_{B,i}$ and $q_{F,i}$,
the potential can still be put in the form (\ref{Pot-gen}),
provided that one substitutes in eqs.(\ref{a-b-c})
$q_B^k \rightarrow \sum_i n_{B,i} q_{B,i}^k$ and
$q_F^k \rightarrow \sum_i n_{F,i} q_{F,i}^k$, where $n_{B(F),i}$ are the multiplicities
of the bosons and fermions with charges $q_{B(F),i}$ ($k=2,3,4$).

For more generic field configurations, such as massive bulk fermions,
localized fields or localized gauge kinetic terms,
the analytical study of the phase transition
is more involved and one has to rely on numerical methods to safely establish
its nature. However, since the coefficient $b(x)$ multiplying the $\alpha^3$ term
in eq.(\ref{Pot-gen}) will still be non-vanishing (unless the model has no massless bosons
coupled to the Wilson line phase),
we expect that the phase transition will be of first order.
We show in next section that this is what generally happens for the model
of  ref.~\cite{Scrucca:2003ra}, by numerically studying the one-loop Higgs potential
in a more complicated setting. We will also see that the analytical studies
performed in this subsection are still approximately valid.

\section{The electroweak phase transition in gauge-Higgs models}

In this section, we want to establish the nature of the electroweak
phase transition in models with gauge-Higgs unification based on
5D orbifolds. We will consider in particular the model constructed in
 ref.~\cite{Scrucca:2003ra} and study if and how the electroweak
phase transition occurs in it and some of its extensions.

\subsection{Review of the model}

The model of  ref.~\cite{Scrucca:2003ra} is based on a 5D gauge theory with gauge group $G=SU(3)_c\times
SU(3)_w$ on an $S^1/\Z_2$ orbifold.\footnote{As explained in detail in  ref.~\cite{Scrucca:2003ra},
an extra $U(1)^\prime$ gauge field has to be introduced in 5D,
in order to get the correct weak mixing angle.
This additional gauge field does not couple with the
Higgs field, and is thus irrelevant in all the considerations that will follow.
For this reason it will be neglected in the following.}
The $\Z_2$ orbifold projection is embedded non-trivially only in the electroweak $SU(3)_w$
gauge group, by means of
the matrix
\begin{equation}
P = e^{i\pi \lambda_3} =
\left(
\matrix{
-1 \a 0 \a 0 \cr
 0 \a -1\;\;\, \a 0 \cr
 0 \a  0 \a 1 \cr}\;
\right)\;,\label{Rtwist}
\end{equation}
where $\lambda_a$ are the $SU(3)$ Gell-Mann matrices,
normalized as ${\rm Tr}\, \lambda_a \lambda_b = 2 \delta_{ab}$.
The twist (\ref{Rtwist}) breaks the electroweak gauge group in 4D to
$SU(2) \times U(1)$.
The massless 4D fields are the gauge bosons $A_\mu^a$ in the
adjoint of $SU(2)$ and a charged scalar doublet $H$, the Higgs field,
arising from the internal components $A_5^a$ of the gauge field.
A VEV for $A_5^a$ induces an additional spontaneous symmetry
breaking to $U(1)_{EM}$. Following  ref.~\cite{Scrucca:2003ra}, we can take
\begin{equation}
\langle A_5^a \rangle = \frac {2\alpha}{g_5 R}\,\delta^{a7} \,,
\label{vev}
\end{equation}
corresponding to a purely imaginary Higgs VEV $\langle H\rangle = 2i\alpha/(g_4 R)$,
where $\alpha$ is a Wilson line phase, $0\leq \alpha \leq 1$, and
$g_4 = g_5/\sqrt{L}$ is the 4D weak coupling constant.

The exact symmetry $\alpha\rightarrow \alpha+1$ prevents
the appearance of any local potential for the Higgs field, at any
order in perturbation theory. The only allowed potential terms
for the Higgs are non-local in the extra dimension (Wilson lines)
and thus free of any divergence \cite{Non-local}. Despite the model is a non-abelian
gauge theory in 5D, and thus non-renormalizable, the Higgs potential
is insensitive to UV physics and thus calculable.
The gauge hierarchy problem is automatically solved in these models.

Introducing matter fields in this set-up is a non-trivial task.
One possibility is to introduce
massive 5D bulk fermions and massless localized chiral fermions, with a mixing
between them, so that the matter fields are identified as the lowest KK mass
eigenstates. In this way, a realistic pattern of Yukawa couplings can
be obtained \cite{Scrucca:2003ra}.\footnote{It should be said, however, that
the construction of an UV completion leading to this fermion spectrum might be
a non-trivial task. Indeed, the latter does not satisfy the requirements
recently found in  ref.~\cite{Serone:2004yn} that allow to consider this model as a degeneration
limit of a well-defined 6D model on $T^2/\Z_2$. We thank A. Wulzer for this
observation.}

The electroweak phase transition occurs in these models because the
radiatively induced one-loop Higgs potential, at zero temperature,
is unstable at $\alpha=0$, in presence of matter fields.
In the broken phase, the mass of the matter fields is exponentially
suppressed in terms of the bulk fermion masses $M_i$. Since the contribution
to the Higgs potential of massive 5D bulk fermions is also exponentially suppressed
in $M_i$, effectively the electroweak phase transition is driven by the
lightest bulk fermion, the one giving rise, through its mixing with the corresponding
localized chiral fermion, to the top quark and top Yukawa coupling.
In this sense, the electroweak phase transition is driven by
the top quark.
We hence focus in the following to the
study of the contribution to the
Higgs effective potential at finite temperature, given by the gauge fields and by
a single bulk fermion, coupled to the boundary fermions which give rise to the top quark.
Subsequently, in order to get more
realistic values for the Higgs and top mass, as well as for the compactification scale
$1/R$, we will consider some extensions of this minimal model, as already
suggested in  ref.~\cite{Scrucca:2003ra}.

The bulk 5D fermion that has the correct quantum number to couple with
the (conjugate) top quark is the symmetric representation ${\bf 6}$ of $SU(3)_w$.\footnote{Thanks
to the extra $U(1)^\prime$ symmetry, it is actually possible to couple
the top quark with larger representations of $SU(3)_w$, resulting in this way
to a larger value for the top mass. We will not pursue this possibility.}
More precisely, the matter fermion of this basic construction is as follows.
We introduce a couple of bulk fermions $\Psi$ and $\tilde\Psi$
with opposite $\Z_2$ parities, in the representation $(\bar{\bf 3},{\bf 6})$
of $SU(3)_c\times SU(3)_w$. At the orbifold fixed points, we have a
left-handed doublet $Q_L=(t_L,b_L)^T$ and a
right-handed fermion singlet $t_R$ of $SU(2)\times U(1)$.
They are located respectively at $y=0$ and $y=L/2$,
the two boundaries of the segment $S^1/\Z_2$.
The parity assignments for the bulk fermions allow for a bulk mass term
$M$ mixing $\Psi$ and $\tilde \Psi$, as well as boundary
couplings $e_{1,2}$ with mass dimension $1/2$ mixing the
bulk fermion $\Psi$ to the boundary fermions $Q_L$ and $t_R$.
The matter Lagrangian reads
\begin{eqnarray}
\mathcal{L}_{\rm mat} = & &
\Big[\bar \Psi \,i \Dslash_5\, \Psi
+ \tilde{\bar \Psi} \,i \Dslash_5\, \tilde \Psi
+ \Big(\bar \Psi M \tilde \Psi +\mathrm{h.c.} \Big)\Big] \nn \\
&&  +\,\delta(y) \Big[\bar Q_L \,i \Dslash_4\, Q_L
+ \Big(e_1 \bar Q_R^c \psi
+ \mathrm{h.c.}\Big) \Big] \nn \\
&& +\,\delta(y-\frac L2 ) \Big[\bar t_R \,i \Dslash_4\, t_R
+ \Big(e_2 \bar t_L^c \chi
+ \mathrm{h.c.}\Big)\Big] \,,\label{Lagferm}
\end{eqnarray}
where $\psi$ and $\chi$ are the doublet and singlet $SU(2)$ components
of the bulk fermion $\Psi$,
$\Dslashn_4$ and $\Dslashn_5$ denote the $4D$ and
$5D$ covariant derivatives so that  $\Dslashn_5 = \Dslashn_4 +
i\gamma_5 D_5$.
The exact spectrum of the bulk-boundary fermion system defined by the
Lagrangian (\ref{Lagferm}) can be determined by solving a complicated
trascendental equation. The lighest state is however massless,
as long as $\alpha=0$. For $\alpha\neq0$, the mass $M_{top}$ of the lightest state,
that we identify with the top quark,
is dynamically generated by non-local Wilson-lines operators,
whose coefficients are exponentially suppressed by the bulk mass term $M$.
The $W$ mass equals $M_W=\alpha/R$, whereas the Higgs mass
is radiatively induced and equals
\begin{equation}
M_H^2(\alpha_{min}) = \bigg(\frac{g_4 R}2\bigg)^2 \,
\frac{\partial^2 V}{\partial \alpha^2}\bigg|_{\alpha=\alpha_{min}} \;,
\label{MH}
\end{equation}
with $V(\alpha)$ the zero temperature Higgs effective potential and $\alpha_{min}$ its minimum
with the lowest value of $\alpha$.
This is typically the global minimum of the potential in the broken phase
and is also the first one appearing during the electroweak phase transition.
Even when $\alpha_{min}$ is only a local minimum, semiclassical analysis
have shown that its tunneling decay rate is negligibly small,
leading to a practically stable minimum \cite{Haba:2002py}.

Finally, let us comment on the range of validity of this theory as an effective
field theory. Using na\"\i ve Dimensional Analysis (NDA), we can estimate the UV cut-off
scale $\Lambda$ to be
given by the scale in which the $5D$ gauge couplings become strong.
Using respectively the strong $SU(3)_c$ and weak $SU(3)_w$ coupling
constants, one gets $\Lambda_c \sim 10/R$ and $\Lambda_w\sim 100/R$ \cite{Scrucca:2003ra}.

\subsection{Results}

We report below our study of the electroweak phase transition
in the model of  ref.~\cite{Scrucca:2003ra} and its generalizations.
The results, including all the figures, have been obtained by
a numerical computation of the exact one-loop relations
reported in the appendix.

\subsubsection{Only bulk fields}

This is the simplest set-up, in which we consider gauge fields and
a couple of bulk fields, with opposite $\Z_2$ parities,
in the representation $({\bf 6},\bar {\bf 3})$ of
$SU(3)_w\times SU(3)_c$. No boundary fermions are introduced.
\begin{figure}[h!]
\begin{center}
\includegraphics[width=.47\textwidth]{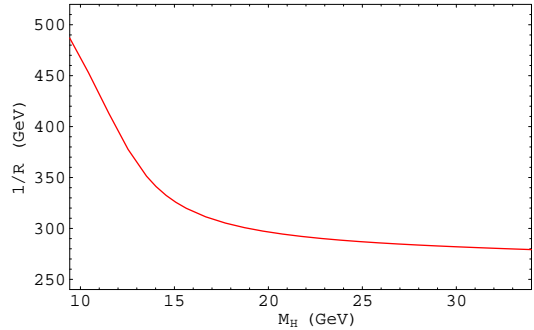}
\caption{Compactification radius as function of the Higgs mass (only bulk fields).}\label{figRadiusSimpl}
\end{center}
\end{figure}

In this theory, a first order phase transition appears for $\lambda \leq 2.1$,
where $\lambda = ML /2$ is the properly rescaled bulk fermion mass,
while outside this range no symmetry breaking occurs. The critical temperature
of the transition is $T_C\sim 1/L$.
\begin{figure}[h!]
\begin{center}
\begin{tabular}{cc}
\includegraphics[width=.47\textwidth]{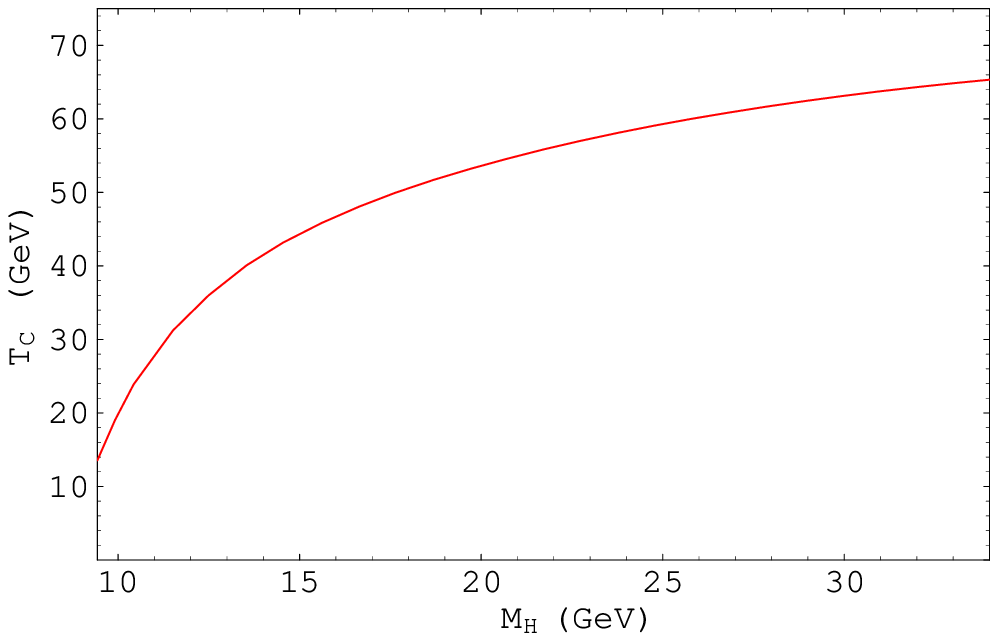}
&\includegraphics[width=.47\textwidth]{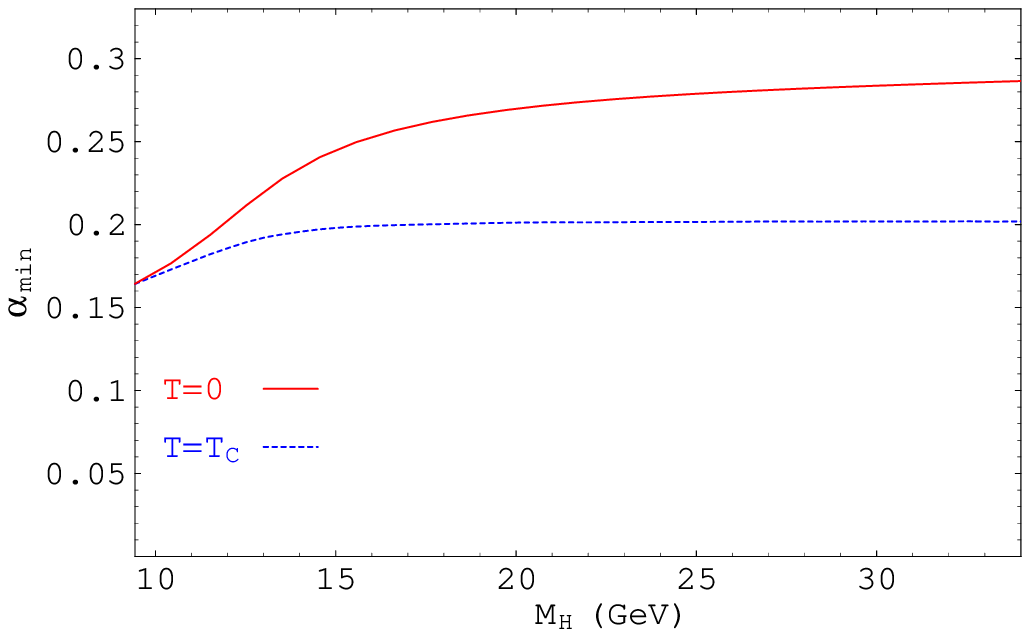}\\
(a)&(b)
\end{tabular}
\caption{(a) Critical temperature and (b) position of the minimum at $T=0$ and at
$T=T_C$, as function of the Higgs mass (only bulk fields).}\label{figCrTempSimpl}
\end{center}
\end{figure}
\begin{figure}[h!]
\begin{center}
\begin{tabular}{cc}
\includegraphics[width=.47\textwidth]{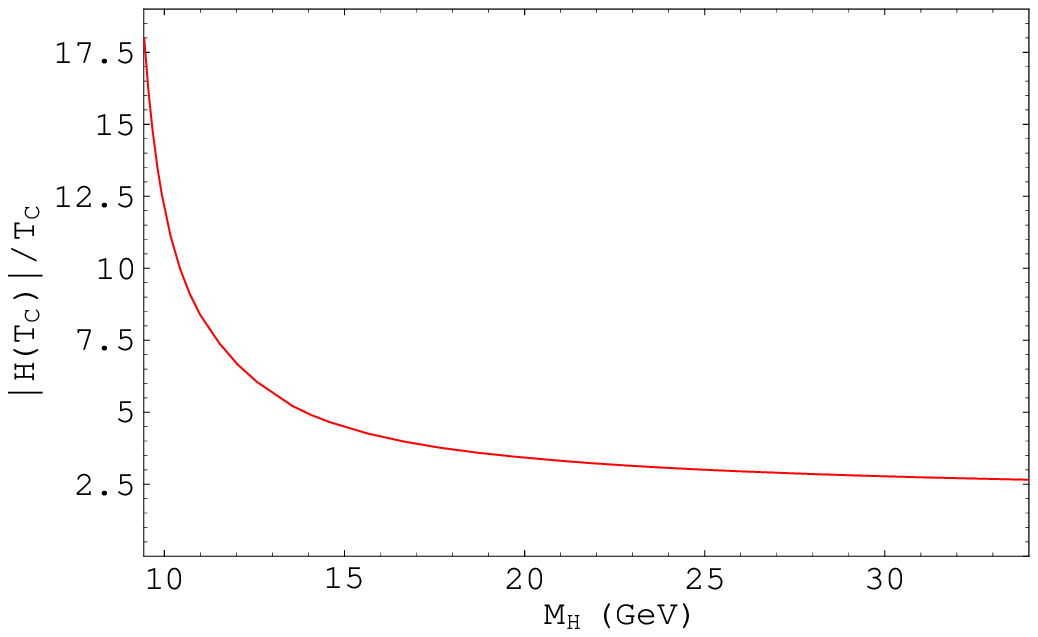}
&\includegraphics[width=.47\textwidth]{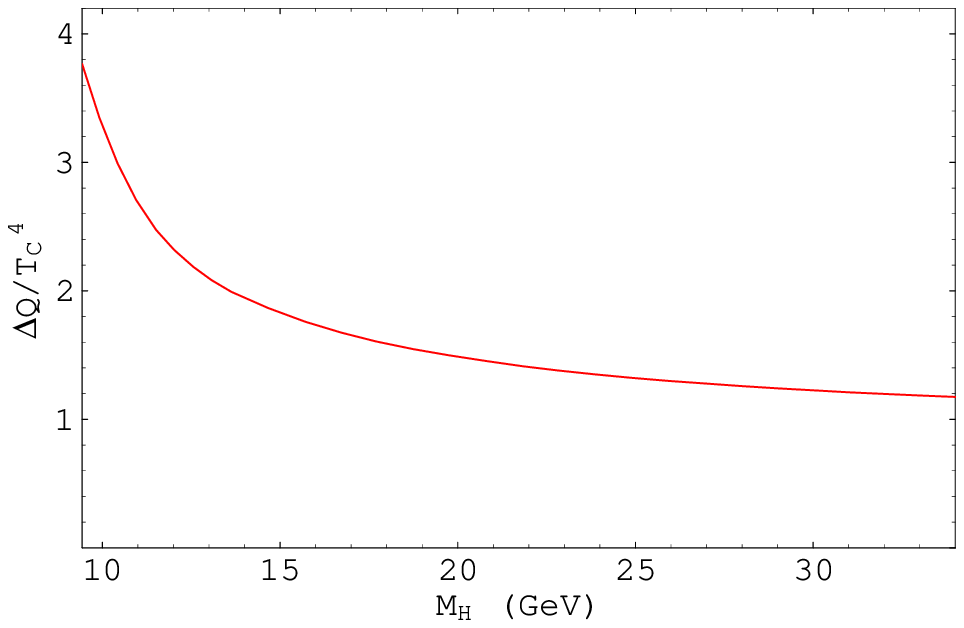}\\
(a)&(b)
\end{tabular}
\caption{(a) Phase transition strength and (b) latent heat as function of the Higgs
mass (only bulk fields).}\label{figPhTrStrSimpl}
\end{center}
\end{figure}
We report in figures \ref{figRadiusSimpl} and \ref{figCrTempSimpl}
how the compactification radius, the critical temperature and
the minimum of the potential (at $T=0$ and at $T=T_C$) depend on the Higgs mass, whose
value depends on $\lambda$.
In figure \ref{figPhTrStrSimpl} (a), we plot
$|H(T_C)|/T_C$ as function of $M_H$.
The phase transition is strongly first order,
as expected for such low values of $M_H$.
In figure \ref{figPhTrStrSimpl} (b), we plot the latent heat as function of $M_H$ (normalized in
units of the critical temperature), whose expression is given by
\be
\Delta Q = \left.T\frac{\partial}{\partial T}V(T,\alpha_{min})\right|_{T=T_C}
-\left.T\frac{\partial}{\partial T}V(T,\alpha=0)\right|_{T=T_C}\,.
\label{deltaQ}
\ee
For simplicity, we normalize the effective potential
so that $V(\alpha=0,T)=0$, in which case the second term in eq.(\ref{deltaQ}) vanishes.
\begin{figure}[h!]
\begin{center}
\begin{tabular}{cc}
\includegraphics[width=.47\textwidth]{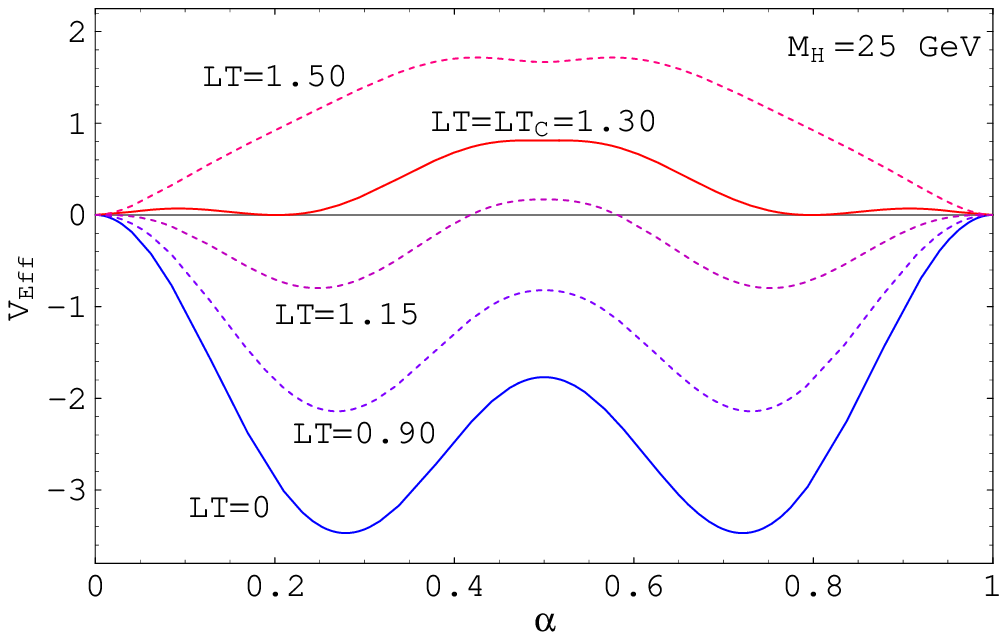}
&\includegraphics[width=.47\textwidth]{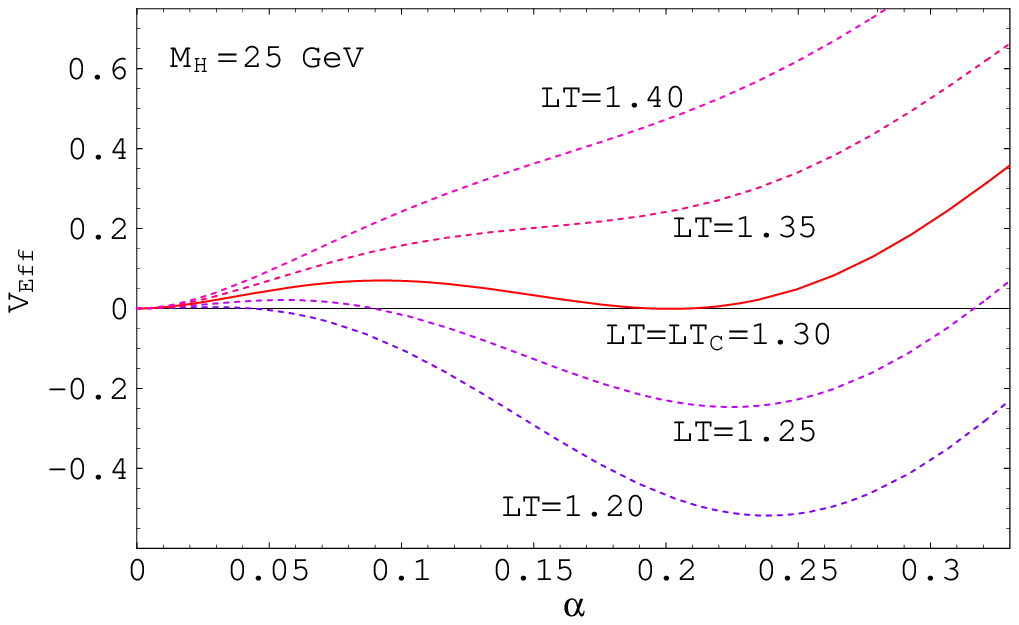}\\
(a) & (b)
\end{tabular}
\caption{(a) Effective potential (for $M_H = 25\, \textrm{GeV}$) at various temperatures.
(b) Detail near the phase transition (only bulk fields).}\label{figEffPotSimpl}
\end{center}
\end{figure}
Finally, in figure \ref{figEffPotSimpl} we show the shape of the effective potential,
for $M_H = 25\, \textrm{GeV}$, at various temperatures.

For high values of $M_H$ (which correspond to low values of the bulk fermion mass $\lambda$),
the behaviour of $\alpha_{min}(T_C)$ and
$|H(T_C)|/T_C$ in figures \ref{figCrTempSimpl} (b) and \ref{figPhTrStrSimpl} (a) can be understood
by applying the considerations of subsection 2.3.
Around the critical temperature the potential (\ref{Pot-gen}) has only a mild
dependence on $\lambda$, at least for $\lambda \lesssim 1.5$, since as first approximation
it implies only the shift $\pi LT\rightarrow \sqrt{4\lambda^2+ (\pi LT)^2}$ in eq.(\ref{Pot-AFer}).
Hence, although the Higgs mass at $T=0$ is sensibly dependent on $\lambda$,
$\alpha_{min}$  and $|H(T_C)|/T_C$, as computed in eqs.(\ref{alfa-min}) and (\ref{H-T}),
are essentially independent of $M_H$ for low values of $\lambda$
(high values of $M_H$).

For larger values of $\lambda$, both $T_C$ and $M_H$
starts to exponentially depend on $\lambda$, resulting in a linear
dependence of $T_C$ on $M_H$.
On the contrary, $\alpha_{min}(T_C)$ remains still constant.
These behaviours are roughly reproduced by the left part
of figures \ref{figCrTempSimpl} (b)  and \ref{figPhTrStrSimpl} (a).

\subsubsection{Inclusion of boundary fermions}\label{secBoundaryFermions}

\begin{figure}[h!]
\begin{center}
\begin{tabular}{cc}
\includegraphics[width=.47\textwidth]{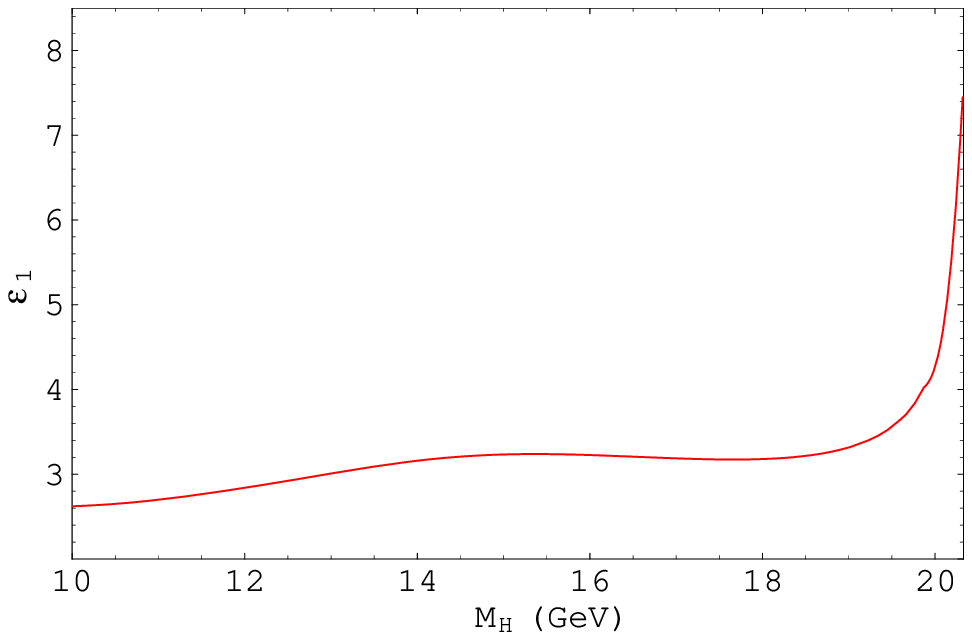}
&\includegraphics[width=.47\textwidth]{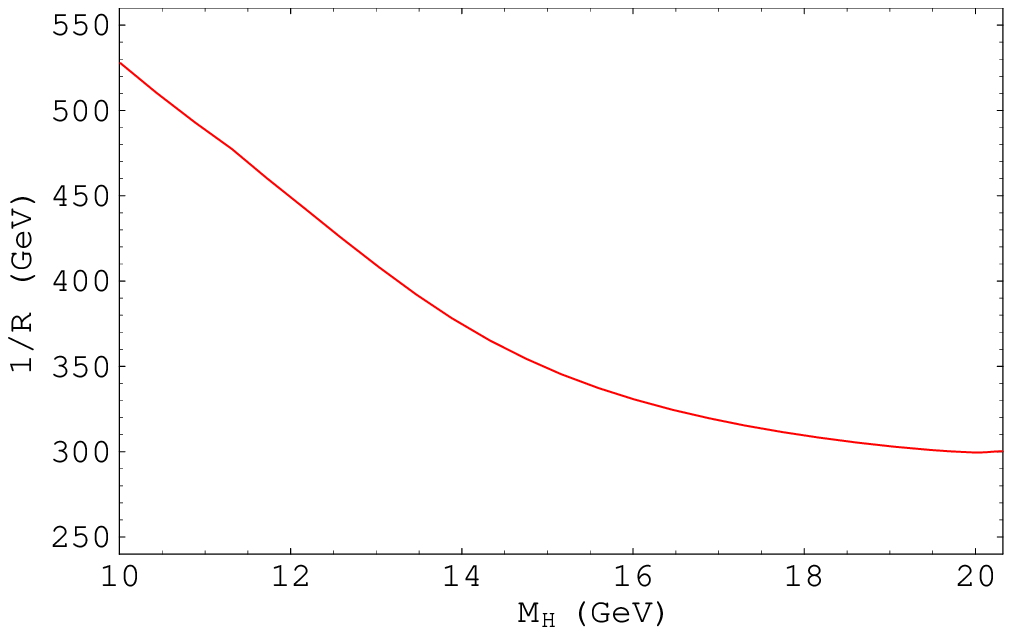}\\
(a) & (b)
\end{tabular}
\caption{(a) Values of the $\epsilon_1$ coupling and (b) compactification radius as function of the Higgs mass
(inclusion of boundary fermions).}\label{figRadiusBFerm1}
\end{center}
\end{figure}

\begin{figure}[h!]
\begin{center}
\begin{tabular}{cc}
\includegraphics[width=.47\textwidth]{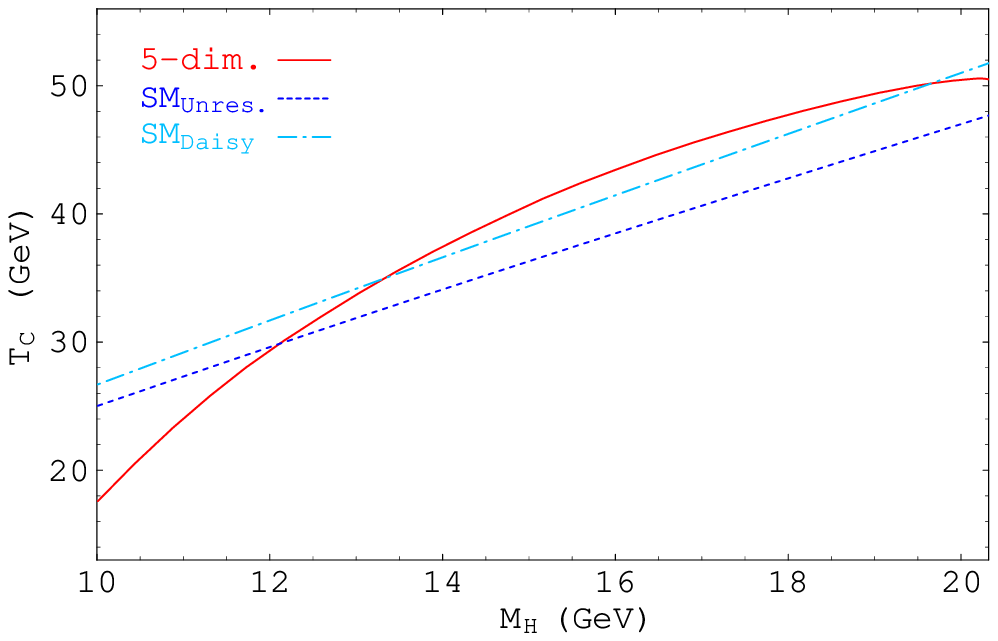}
&\includegraphics[width=.47\textwidth]{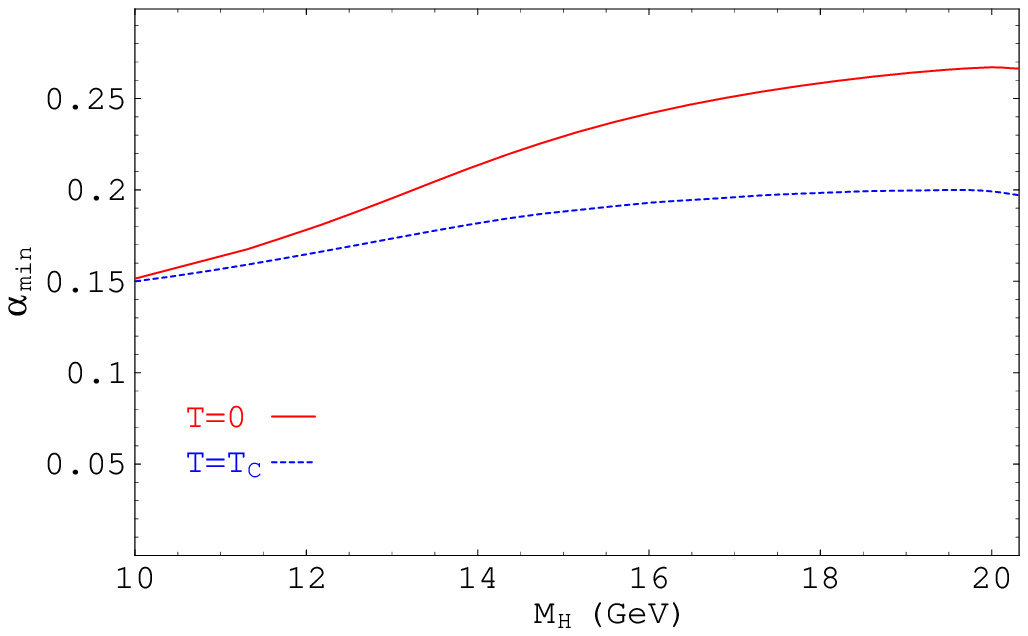}\\
(a) & (b)
\end{tabular}
\caption{(a) Critical temperature and (b) position of the minimum at $T=0$ and at $T=T_C$,
as function of the Higgs mass (inclusion of boundary fermions).}\label{figCrTempBFerm1}
\end{center}
\end{figure}

\begin{figure}[h!]
\begin{center}
\begin{tabular}{cc}
\includegraphics[width=.47\textwidth]{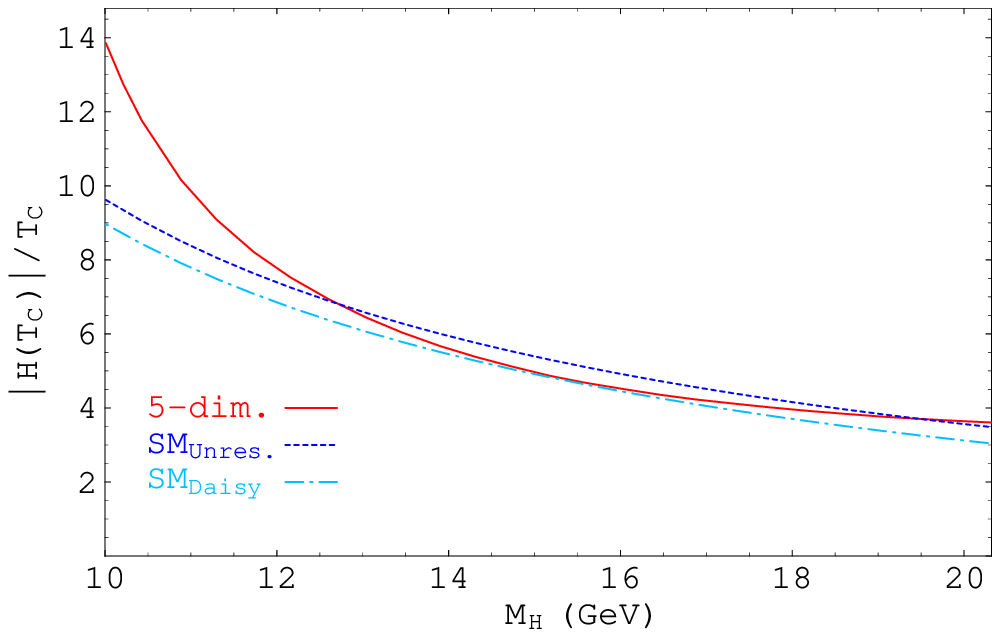}
&\includegraphics[width=.47\textwidth]{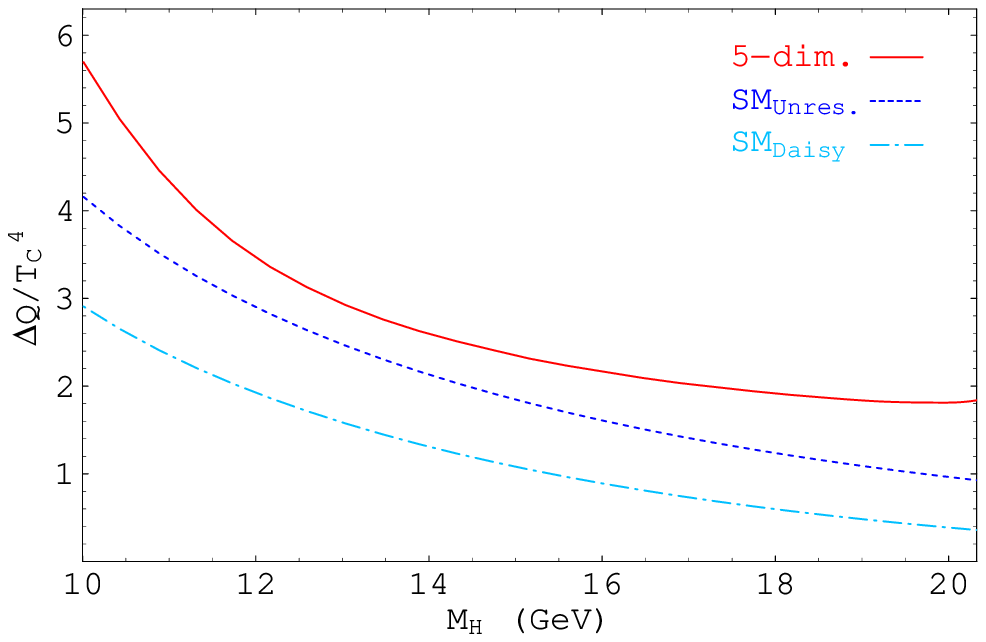}\\
(a) & (b)
\end{tabular}
\caption{(a) Phase transition strength and (b) latent heat as function of the Higgs
mass (inclusion of boundary fermions).}\label{figPhTrStrBFerm1}
\end{center}
\end{figure}

\begin{figure}[h!]
\begin{center}
\begin{tabular}{cc}
\includegraphics[width=.47\textwidth]{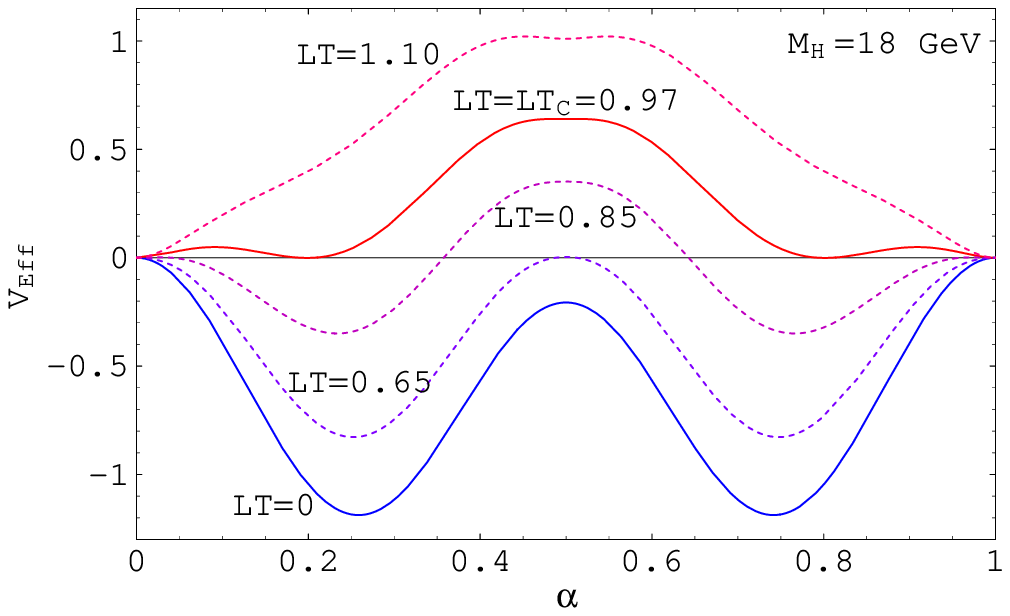}
&\includegraphics[width=.47\textwidth]{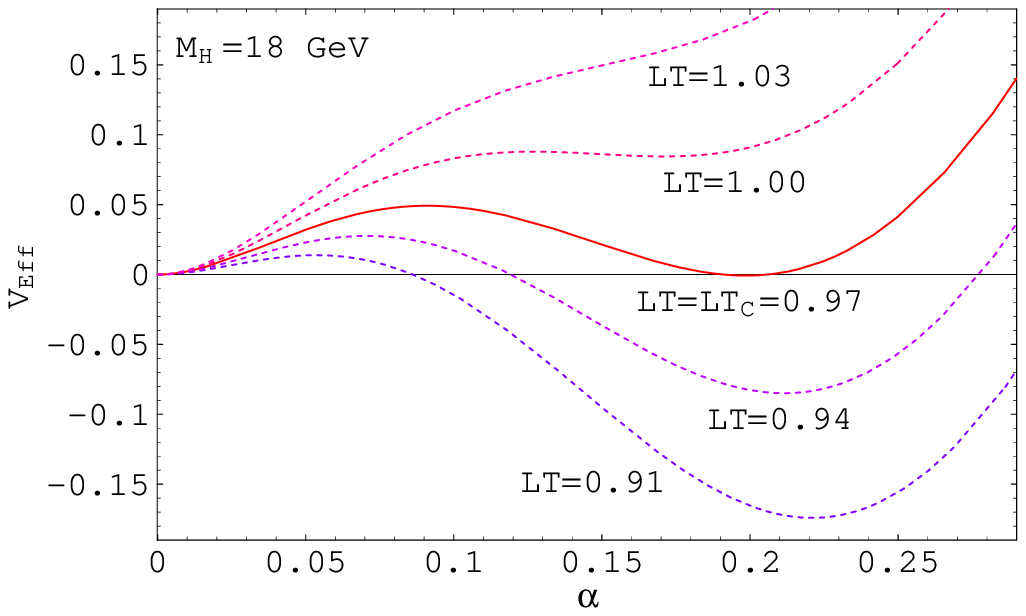}\\
(a) & (b)
\end{tabular}
\caption{(a) Effective potential (for $M_H = 18\, \textrm{GeV}$) at various temperatures. (b) Detail near the phase
transition (inclusion of boundary fermions).}\label{figEffPotBFerm1}
\end{center}
\end{figure}

We now add the boundary fermions, namely the doublet $Q_L$ and the
singlet $t_R$, as in eq.(\ref{Lagferm}).
The relevant formulae for the effective potential in presence of boundary
fields are found in the appendix, eqs.(\ref{Vboundary}).
Since the potential does not significantly depend on the bulk-to-boundary coupling $\epsilon_i$, the
crucial dependence being given by the bulk fermion mass $\lambda$, we have
set $\epsilon_2= \epsilon_1$ and fixed the overall scale by the requirement
of having $M_{top}=45 \,\mathrm{GeV}$.
The effective potential is studied
for $0.85 \leq \lambda \leq 1.85$. Again, the electroweak phase transition is of first order,
with a critical temperature of order $1/L$ (see Figs.~\ref{figRadiusBFerm1}-\ref{figEffPotBFerm1}).


For comparison, in figures \ref{figCrTempBFerm1} (a) and \ref{figPhTrStrBFerm1},
we plotted the critical temperature, $|H(T_C)|/T_C$ and the latent heat as function
of the Higgs mass for both the 5D model and the SM, with $M_{top}=45\, \textrm{GeV}$.
For the SM case, we plot both the na\"\i ve one-loop ($SM_{Unres.}$)
and one-loop improved ($SM_{Daisy}$) results, the latter
obtained by resumming the leading
daisy diagrams \cite{Carrington:1991hz,Arnold:1992rz}.\footnote{We have
fixed the Landau gauge, which turns out to be a good gauge choice.
Indeed, results obtained in
this gauge are very similar to the ones obtained in gauge-invariant approaches to the
electroweak phase transition, such as the gauge-invariant effective potential
\cite{Hebecker:1994} and lattice computations \cite{Lattice}.
We also used a cut-off renormalization
scheme choosing the counterterms so that the position
of the zero temperature minimum of the effective potential and the Higgs mass do not change
with respect to their tree level values \cite{Anderson:1991zb}.}
For such low values of the Higgs mass, $10 \div 20$ GeV,
the SM effective potential is well described by its one-loop approximation.

Even when boundary fermions are inserted, the behaviour of $\alpha_{min}(T_C)$
and $|H(T_C)|/T_C$ reported in Figures \ref{figCrTempBFerm1} (b) and \ref{figPhTrStrBFerm1} (a)
is very similar to that found in the previous case with only bulk fermions.
This is explained by recalling, as already mentioned, that the
potential has a small dependence on the bulk-to-boundary couplings $\epsilon_i$.
All the analytical considerations performed at the end of section 3.2.1
are then approximately valid also in this case.

\subsubsection{Theory with high rank bulk fermions}

One of the main problems of models based on gauge-Higgs unification is the predicted too low value for the Higgs mass.
A possibility to solve this problem is given by the introduction of additional bulk fermions, which do not
couple to the localized matter fields, in high rank representations of $SU(3)_w$ \cite{Scrucca:2003ra}.
\begin{figure}[h!]
\begin{center}
\begin{tabular}{cc}
\includegraphics[width=.47\textwidth]{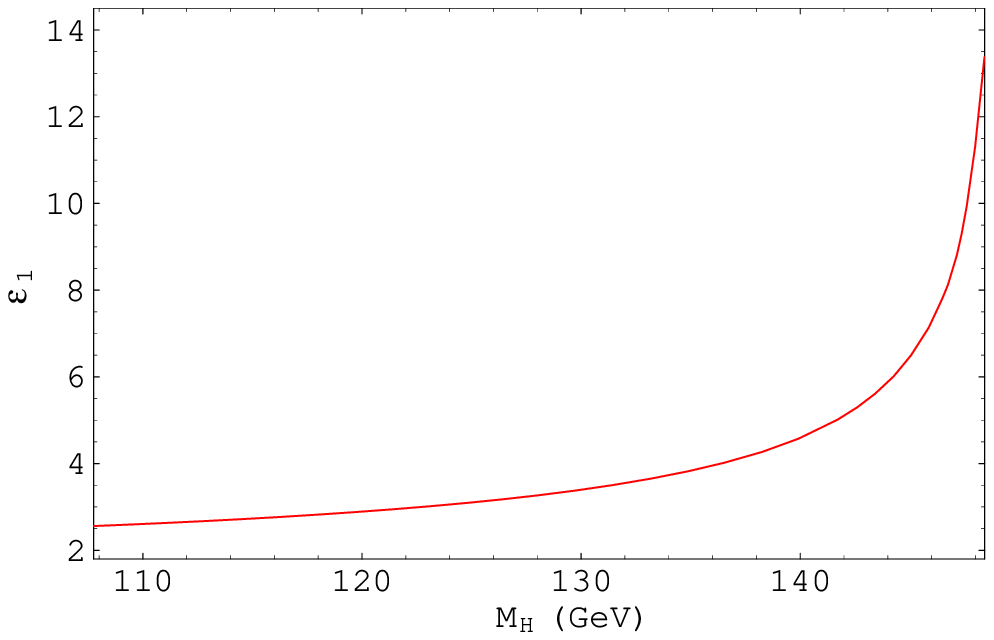}
&\includegraphics[width=.47\textwidth]{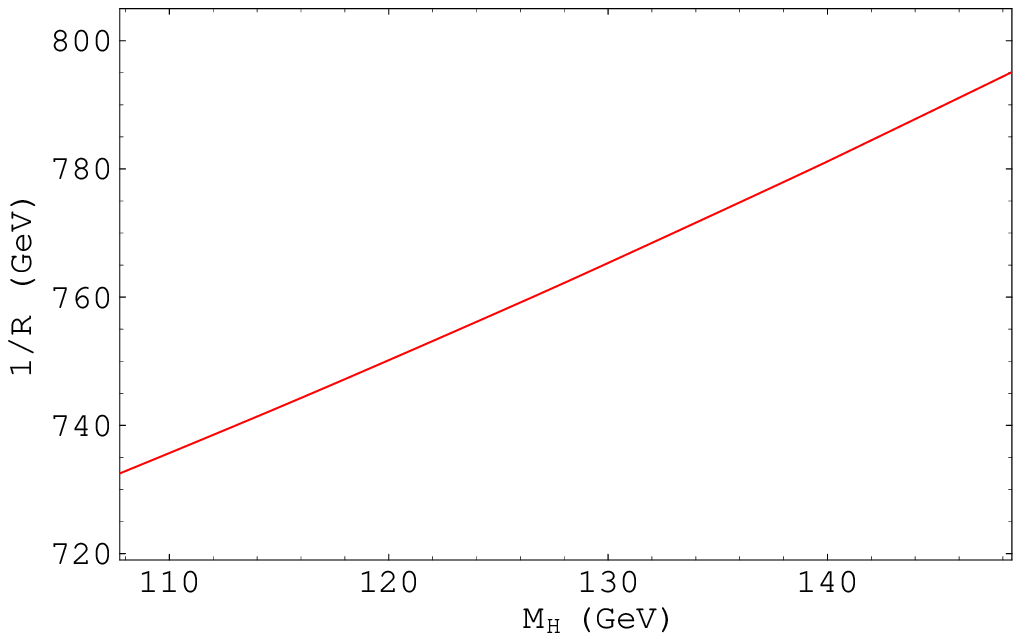}\\
(a) & (b)
\end{tabular}
\caption{(a) Value of the $\epsilon_1$ coupling and (b) compactification radius, as function of the Higgs mass
(high rank bulk fermions).}\label{figRadiusRank8}
\end{center}
\end{figure}
We consider in the following the electroweak phase transition that one obtains in a 5D model where,
in addition to the fields introduced in section 3.2.2, one considers a massive bulk fermion in the
completely symmetric rank $8$ representation of $SU(3)_w$ and singlet of $SU(3)_c$.
The mass of this fermion has been fixed to $\lambda_{HR} = 2.15$.
The bulk-to-boundary couplings $\epsilon_i$ have been fixed requiring that $M_{top}=45\,\mathrm{GeV}$, with
$\epsilon_2 = 0.5\, \epsilon_1$.
The effective potential is studied for $1.6 \leq \lambda \leq 2.4$.
For $\lambda > 1.79$ (which corresponds to $M_H > 125\, \textrm{GeV}$), $\alpha_{min}$
is the global minimum of the potential, which turns to a local minimum for smaller values of $\lambda$.
Our results are summarized in figures \ref{figRadiusRank8}-\ref{figEffPotRank8}.
\begin{figure}[h!]
\begin{center}
\begin{tabular}{cc}
\includegraphics[width=.47\textwidth]{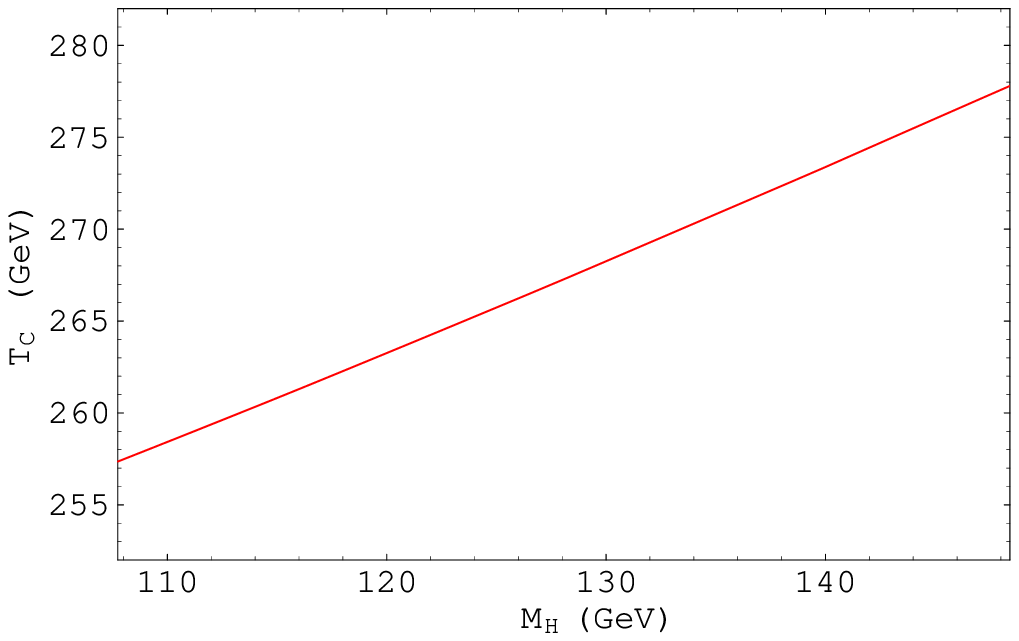}
&\includegraphics[width=.47\textwidth]{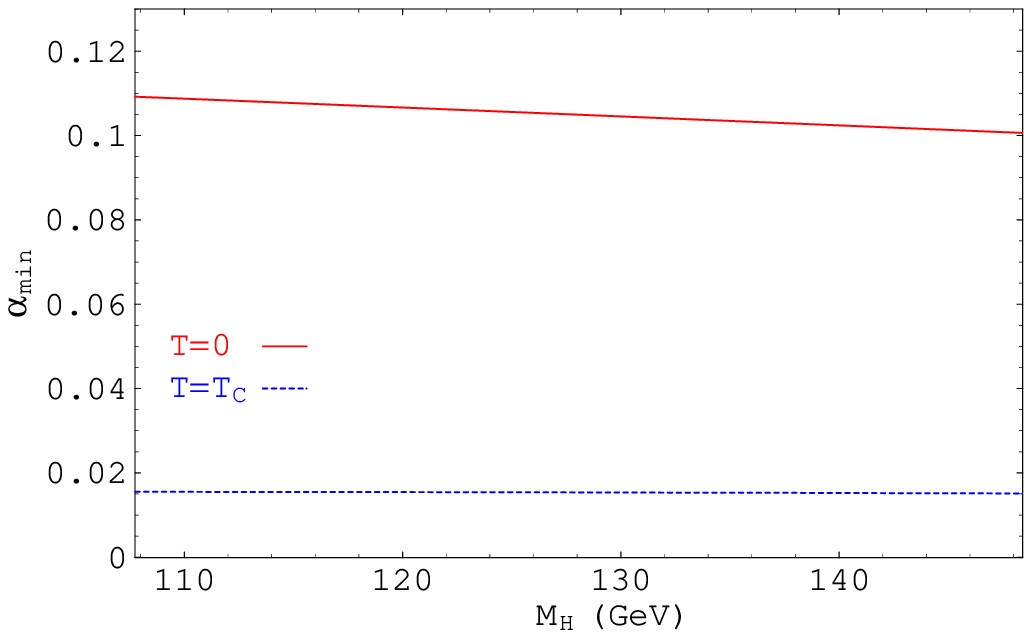}\\
(a) & (b)
\end{tabular}
\caption{(a) Critical temperature and (b) position of the minimum at $T=0$ and at $T=T_C$,
as function of the Higgs mass (high rank bulk fermions).}\label{figCrTempRank8}
\end{center}
\end{figure}
\begin{figure}[h!]
\begin{center}
\begin{tabular}{cc}
\includegraphics[width=.47\textwidth]{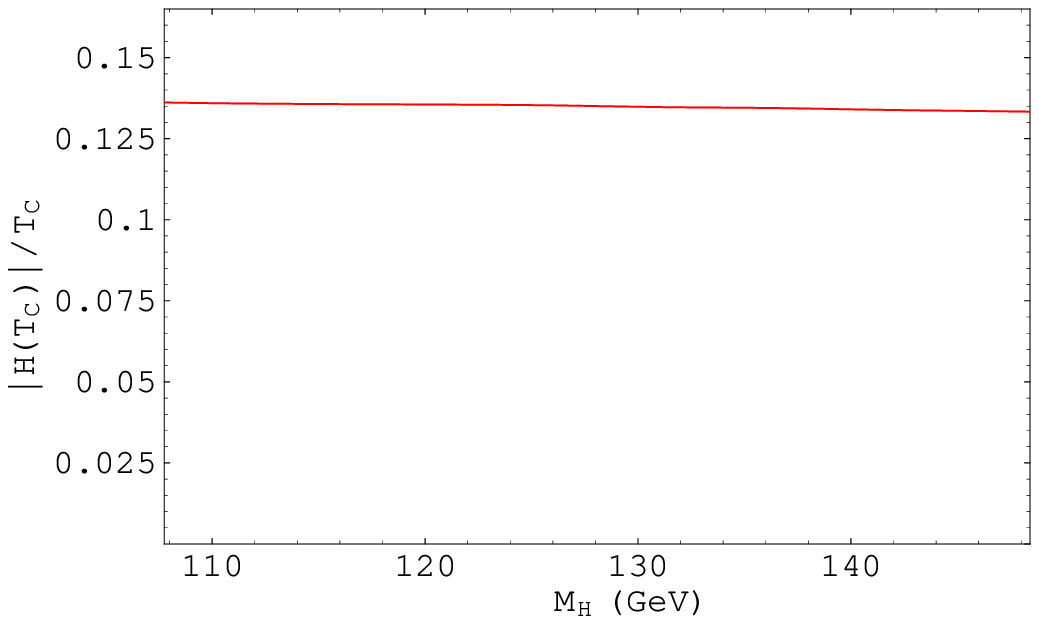}
&\includegraphics[width=.47\textwidth]{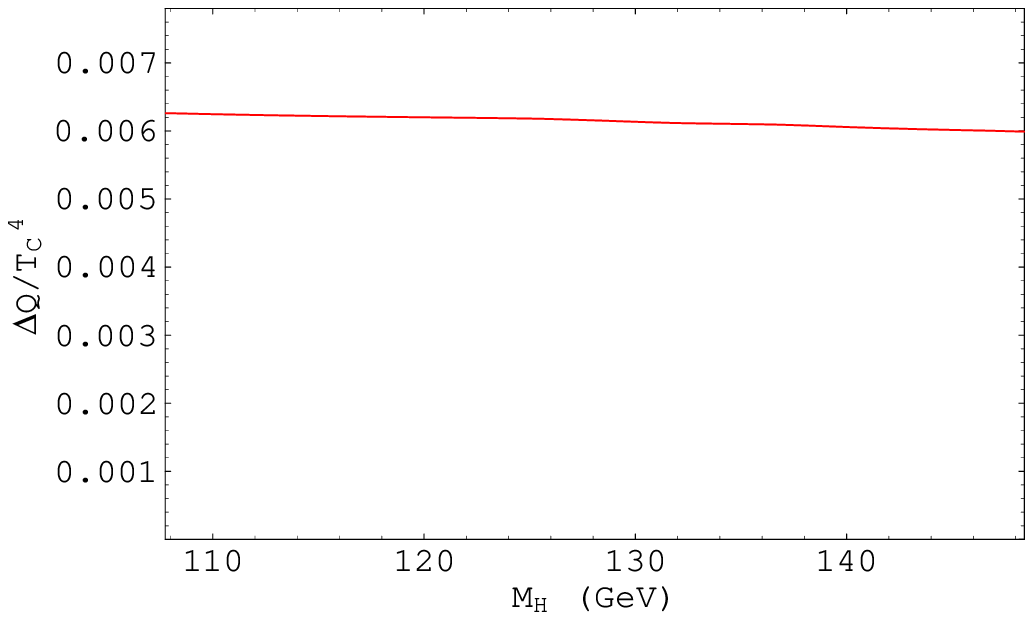}\\
(a) & (b)
\end{tabular}
\caption{(a) Phase transition strength and (b) latent heat as function of the Higgs mass (high rank bulk fermions).}\label{figPhTrStrRank8}
\end{center}
\end{figure}
The Higgs mass can now reach realistic values (although
the top mass is still too low).
As is clear from figure \ref{figPhTrStrRank8} (a), a very weak first order phase transition
is found in this model with $T_C\sim 1/L$.

This behaviour can be understood by means of our general analysis of section 2.3.
The fermionic potential is dominated by the high rank fermion whose mass is fixed. Since
$M_H$ varies only with $\lambda$, the small dependence
on $M_H$ in figures \ref{figCrTempRank8} (b) and \ref{figPhTrStrRank8} is easily understood.
The decreasing of the strength of the phase transition with the rank is clear
by noting that
\be
\alpha_{min}(T_C)\sim \frac{\sum_i n_i q_{i}^2}{\sum_i n_i q_{i}^4}\,,
\ee
where $i$ runs over all the fermion components with charge $q_{i}$ and multiplicity $n_i$,
arising from the decomposition of the rank $r$ bulk fermion.

Fermions in too high-rank representations can not be considered, since the value of the
UV cut-off $\Lambda_w$ rapidly decreases as $r$ increases,
restricting the range of validity of the effective theory \cite{Scrucca:2003ra}.
No comparison with the SM is given, since for such values of the Higgs mass,
perturbation theory breaks down close to the critical temperature.

\begin{figure}[h!]
\begin{center}
\begin{tabular}{cc}
\includegraphics[width=.47\textwidth]{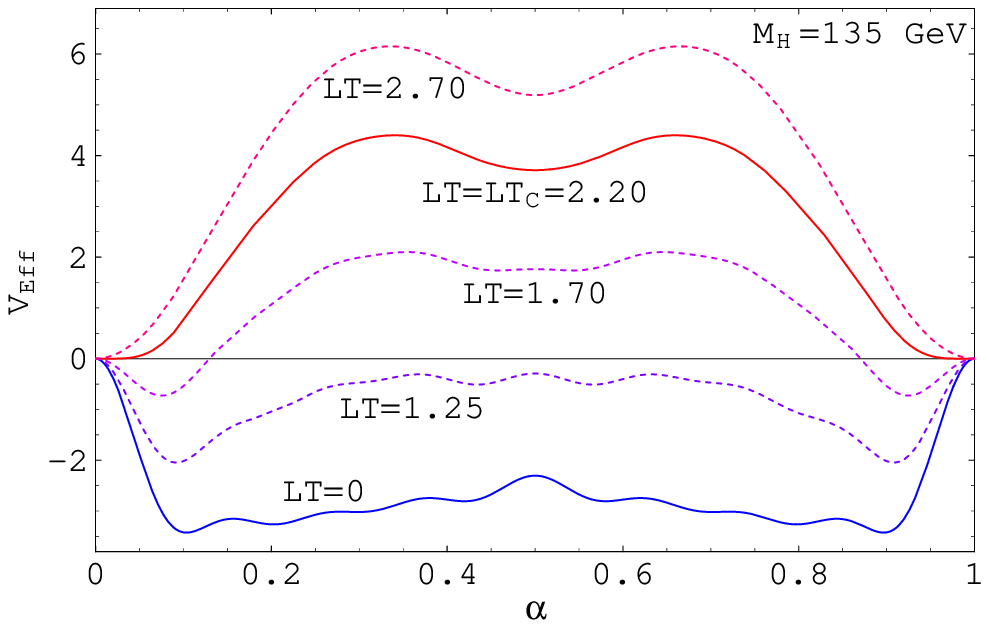}
&\includegraphics[width=.47\textwidth]{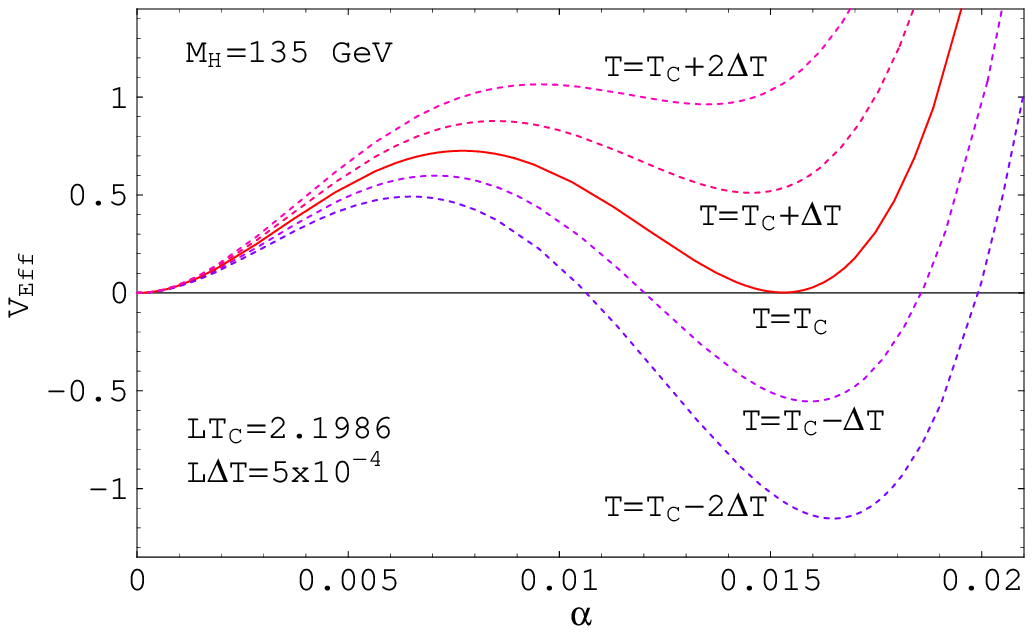}\\
(a) & (b)
\end{tabular}
\caption{(a) Effective potential (for $M_H = 135\, \textrm{GeV}$) at various temperatures. (b)
Detail near the phase transition. The effective potential is multiplied by $10^4$ with respect
to the (a) diagram (high rank bulk fermions).}\label{figEffPotRank8}
\end{center}
\end{figure}

\subsubsection{Theory with localized gauge kinetic terms}

An interesting possibility that is allowed by symmetries in our 5D model is the
introduction of localized gauge kinetic terms. These have a significant effect
on the phenomenology of models based on gauge-Higgs unification.
In particular, they represent another option, in alternative to the introduction of fermions
in high rank representations, to get realistic values for the Higgs mass.
Interestingly enough, their presence also allow to get acceptable values for the top mass.
As a drawback, they break the custodial symmetry, leading to too large deviations
for the $\rho$ parameter.
\begin{figure}[h!]
\begin{center}
\begin{tabular}{cc}
\includegraphics[width=.47\textwidth]{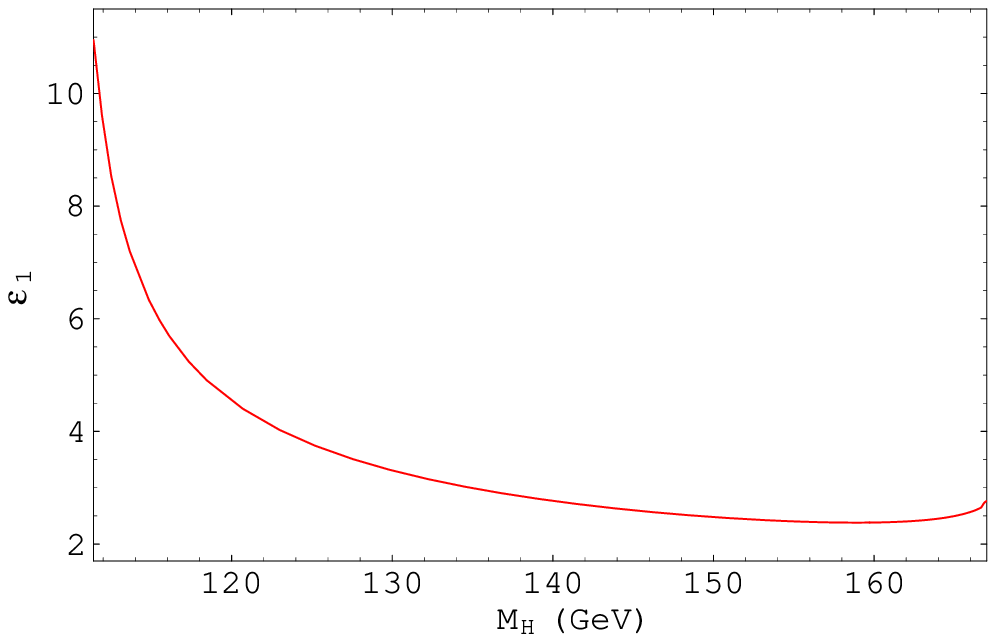}
&\includegraphics[width=.47\textwidth]{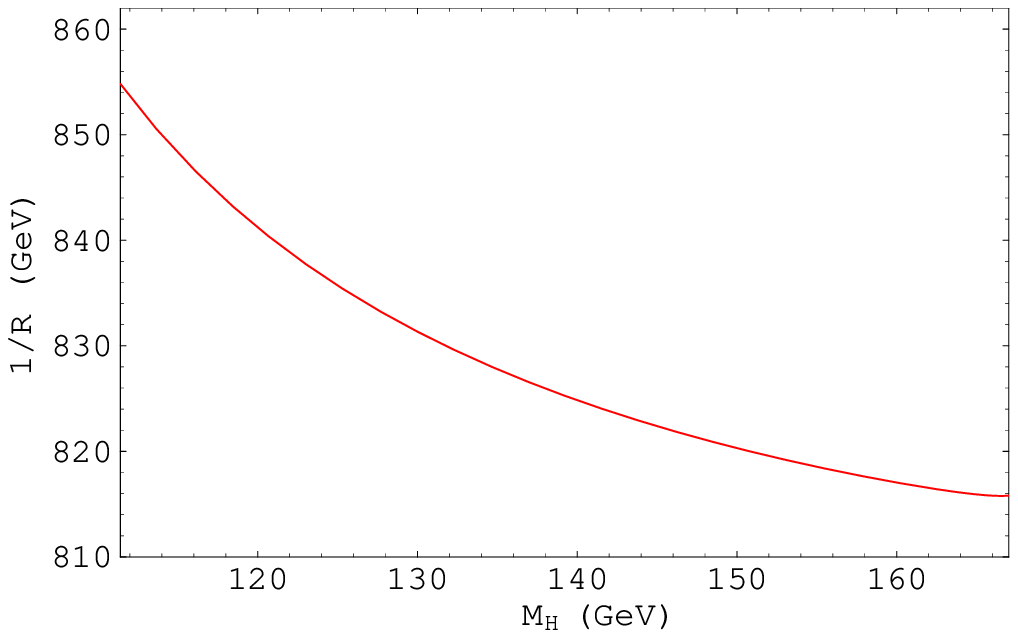}\\
(a) & (b)
\end{tabular}
\caption{(a) Value of the $\epsilon_1$ coupling and (b) compactification radius as function
of the Higgs mass (localized gauge kinetic terms).}\label{figRadiusKin1}
\end{center}
\end{figure}

\begin{figure}[h!]
\begin{center}
\begin{tabular}{cc}
\includegraphics[width=.47\textwidth]{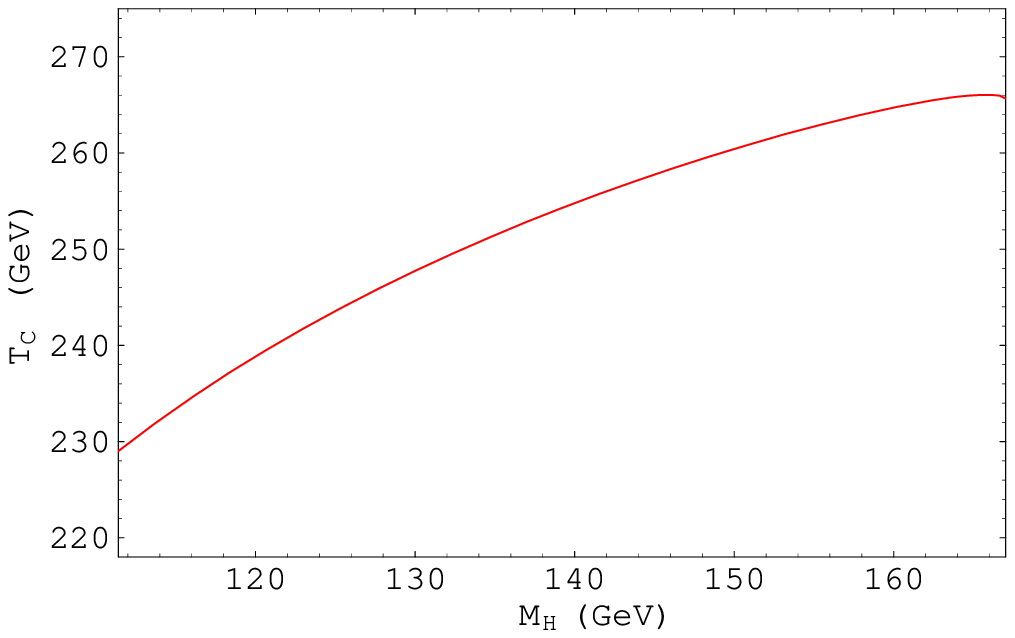}
&\includegraphics[width=.47\textwidth]{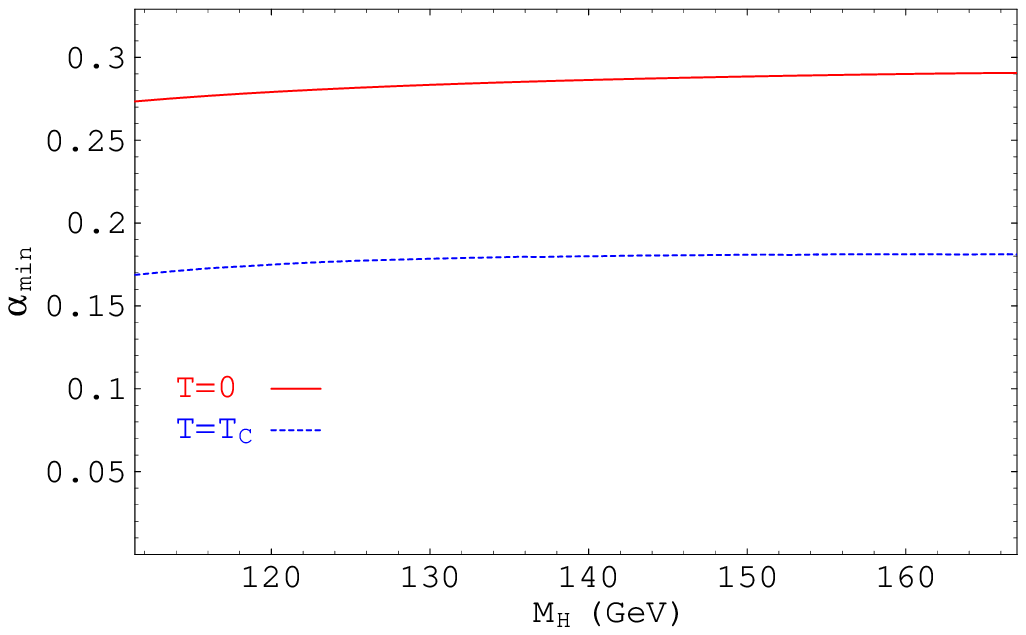}\\
(a) & (b)
\end{tabular}
\caption{(a) Critical temperature and (b) position of the minimum at $T=0$ and at $T=T_C$,
as function of the Higgs mass (localized gauge kinetic terms).}\label{figCrTempKin1}
\end{center}
\end{figure}

\begin{figure}[h!]
\begin{center}
\begin{tabular}{cc}
\includegraphics[width=.47\textwidth]{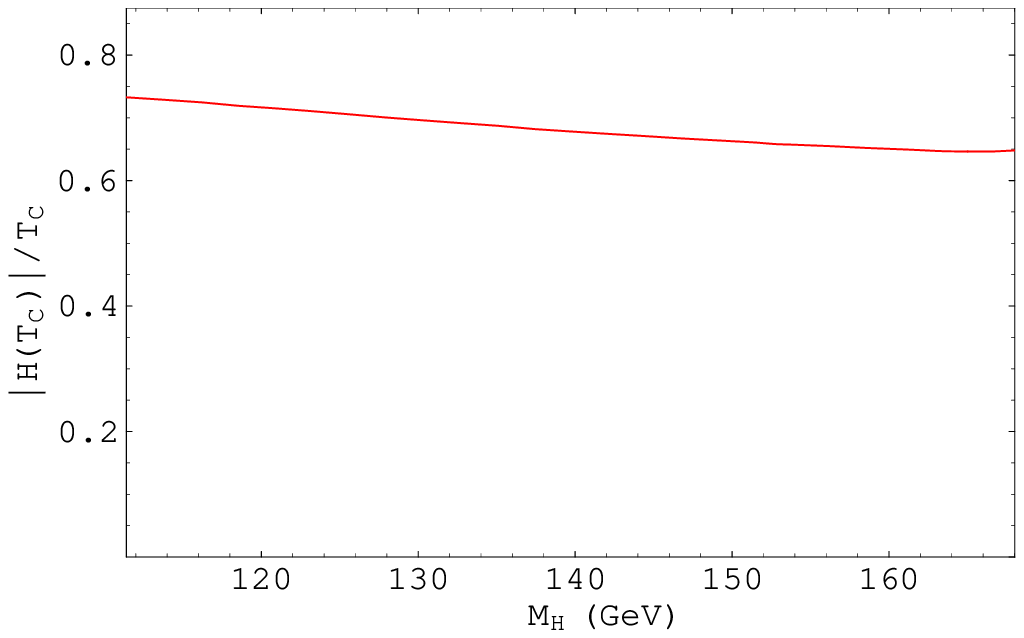}
&\includegraphics[width=.47\textwidth]{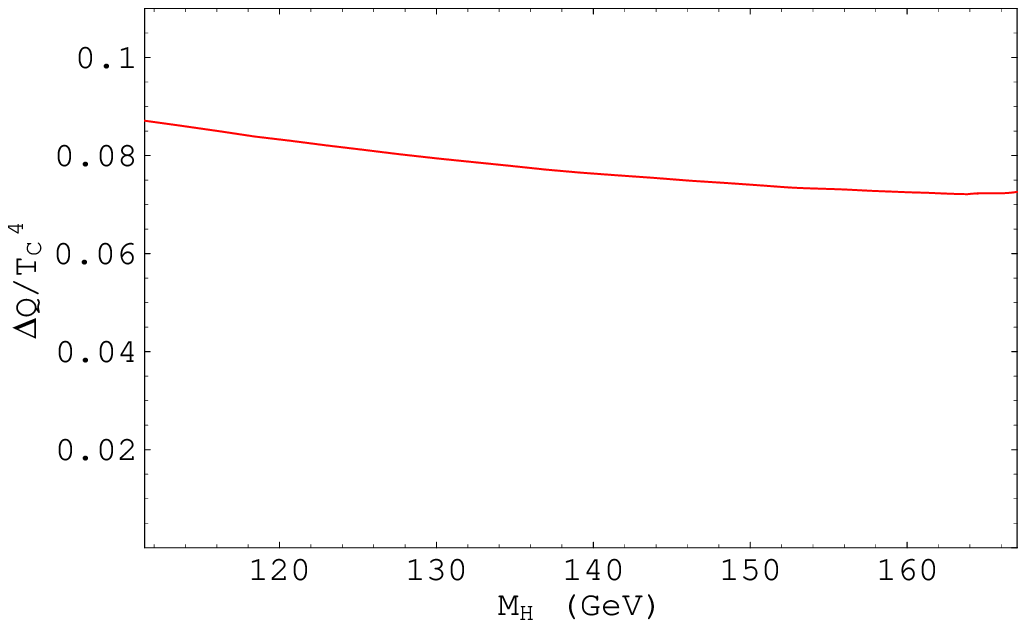}\\
(a) & (b)
\end{tabular}
\caption{(a) Phase transition strength and (b) latent heat as function of the Higgs mass
(localized gauge kinetic terms).}\label{figPhTrStrKin1}
\end{center}
\end{figure}

\begin{figure}[h!]
\begin{center}
\begin{tabular}{cc}
\includegraphics[width=.47\textwidth]{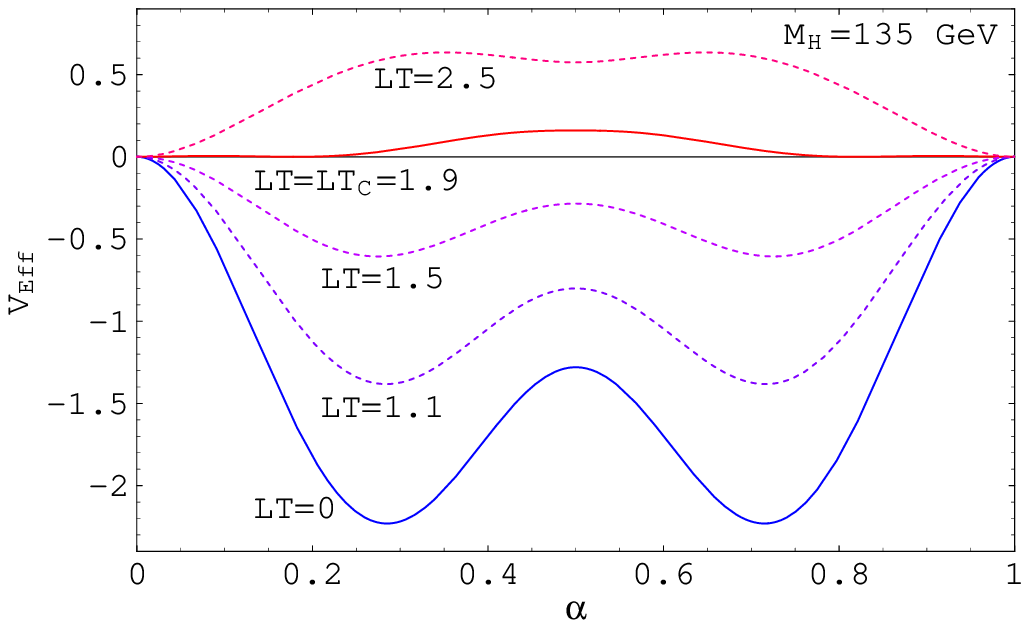}
& \includegraphics[width=.47\textwidth]{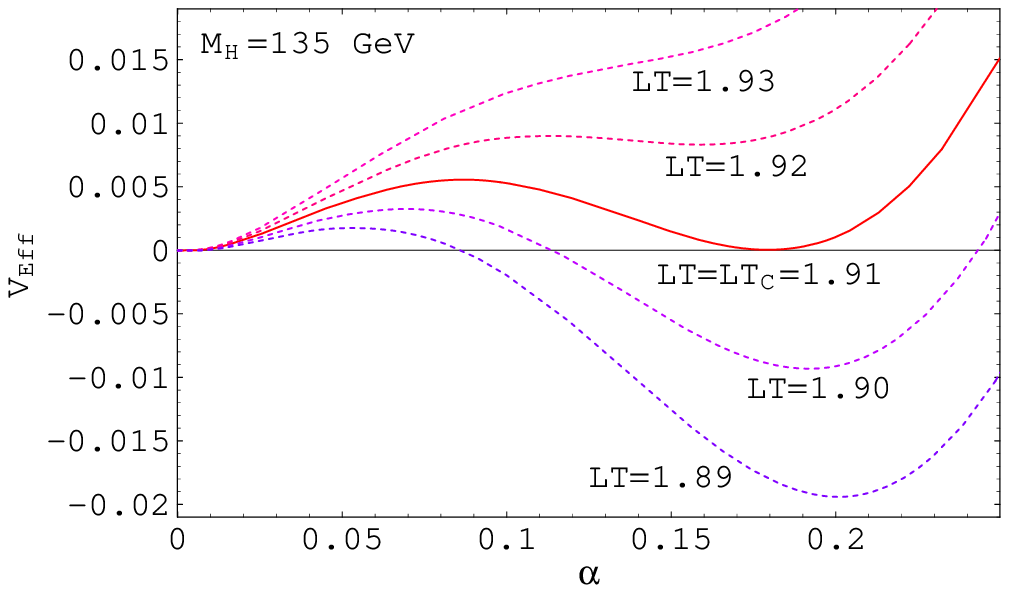}\\
(a) & (b)
\end{tabular}
\caption{(a) Effective potential (for $M_H = 135\, \textrm{GeV}$) at various temperatures. (b) Detail near
the phase transition (localized gauge kinetic terms).}\label{figEffPotKin1}
\end{center}
\end{figure}

We introduce localized gauge kinetic terms at only one fixed point of $S^1/\Z_2$, since
this is the most phenomenological convenient option.
In the notation of  ref.~\cite{Scrucca:2003ra},
we take $c_1 \equiv c = 6$ and $c_2 = 0$.\footnote{When localized gauge kinetic terms are inserted,
the cut-off $\Lambda_w\rightarrow \Lambda_w/c$, so the value chosen for $c$
is well below the limit of validity of the effective theory.}
The gauge contribution to the Higgs effective potential, in presence of
localized kinetic terms, is reported in eq.(\ref{V-BKT}) of the appendix.
The fermions are as in section~\ref{secBoundaryFermions}.
The bulk-to-boundary couplings $\epsilon_i$ are chosen so that the top quark mass is $110\ \mathrm{GeV}$, with
$\epsilon_2 = \epsilon_1$. The effective potential is studied for $0.8 \leq \lambda \leq 1.9$.
Interestingly enough, despite the high value of the Higgs mass obtained in this case,
$110 \div 170$ GeV (see figures \ref{figRadiusKin1}
and \ref{figCrTempKin1}), the phase transition is still of first order and is also moderately strong,
with $|H(T_C)|/T_C\sim 0.7$ (see figure \ref{figPhTrStrKin1} (a)). Since this is of order one,
it might be strong enough to avoid that sphalerons in the broken phase wash out the previously
generated baryon asymmetry (see footnote \ref{footBaryonAsymmetry}).
The critical temperature in this model is $\sim 2/L$ (see figures \ref{figRadiusKin1} (b) and
\ref{figCrTempKin1} (a)).

The behaviour of the phase transition for large values of $c$ is analytically harder to be studied
than the previous cases, since the localized gauge field contribution (\ref{V-BKT})
plays a crucial role. By  expanding the total gauge field contribution, we have been
able to establish that the first two terms of the potential in an $\alpha$ expansion,
namely the quadratic and cubic term in eq.(\ref{Pot-gen}), scales like $a(x)\sim 1/c$ and
$b(x)\sim c^{-3/2}$.  We have not been able to find the asymptotic behaviour for large $c$
of the quartic coupling, which seems to be negative. Neglecting the quartic coupling arising
from the bosonic potential, we find that the critical temperature depends only logarithmically on $c$
and $\alpha_{min}(T_C)\sim 1/\sqrt{c}$. Since $g_5\sim g_4\sqrt{L c}$, we get $H(T_C)/T_C\sim 1/c$.
Our numerical analysis confirm the decrease of the phase transition strenght with $c$, but
numerically we find $|H(T_C)|/T_C\sim 1/c^{1/2\div 1}$, which indicates that the bosonic quartic
coupling cannot be totally neglected and result in a slightly stronger first-order phase transition.

\section{Estimate of higher-loop corrections}

All the analysis performed in this paper about the electroweak
phase transition is based on the computation of a
one-loop effective potential.
Aim of this section is to
estimate the leading higher loop contributions
and determine the range of validity of our one-loop analysis.

\begin{figure}[h!]
\begin{center}
\begin{tabular}{c@{\hspace{4em}}c}
\includegraphics[width=.25\textwidth]{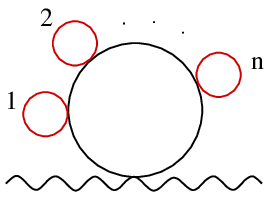}
&\includegraphics[width=.3\textwidth]{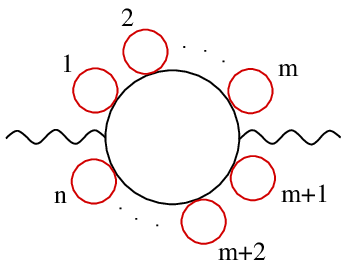}\\
(a) & (b)
\end{tabular}
\caption{Examples of higher loop daisy diagrams contributing to the mass of the
Wilson line phase. Solid and wavy lines represent respectively the charged scalar
and the Wilson line phase. For simplicity, we have omitted the arrows in the
charged scalar propagators.}\label{figGaugeDaisy}
\end{center}
\end{figure}

It is known that the most important higher loop contributions to the
effective potential in 4D theories generally arise from certain IR divergent diagrams,
known as ``daisy'' or ``ring'' diagrams \cite{Dolan:1973qd} (see figure \ref{figGaugeDaisy}).

As shown in ref.~\cite{Parwani:1991gq}, the latter are all easily resummed by adding and subtracting
to the Lagrangian the one-loop (thermal) mass correction for any field present in the Lagrangian.
One then defines a new propagator by adding the thermal mass to the tree level one, and consider
the other (opposite)
term as a one-loop mass insertion. Such procedure has been performed both at one
\cite{Carrington:1991hz} and two-loop level \cite{Arnold:1992rz} in the SM.
Interestingly enough, the two-loop improved SM effective potential
turns out to give results very similar to that
computed by numerical results based on lattice simulations even for values of $M_H$
close to (but less than) $M_W$, in which case perturbation theory is expected to be a bad approximation.

In the following, we will give an estimate of the contribution of daisy diagrams for models
where the order parameter is a Wilson line phase, as in theories with gauge-Higgs unification.
For simplicity, we focus on a particularly simple model, scalar QED in 5 dimensions, compactified
on a circle $S^1$ of length $L=2\pi R$.\footnote{We take $S^1$ and not $S^1/\Z_2$ as compact space
because in the former case we can study an abelian gauge field model, whereas in the latter
case the $\Z_2$ orbifold parity forbids the presence of abelian Wilson line phases
and would thus oblige us to consider non-abelian gauge fields.}
The Lagrangian is
\be
{\cal L} = (D_M \Phi)^\dagger D^M \Phi  - \frac{\lambda_5}{4}(\Phi^\dagger \Phi)^2
- \frac 14 F_{MN}F^{MN}\,,
\label{lag-Phi}
\ee
where $\Phi$ is a complex scalar field with periodic boundary conditions: $\Phi(2\pi R) = \Phi(0)$.
Our aim is to study the effective potential for the Wilson line phase $\alpha=A_5 g_5 R$,
in the approximation in which the dimensionless 4D gauge coupling
$g_4\ll \lambda$, with $\lambda = \lambda_5/L$.\footnote{By performing
a non-single valued (on $S^1$) gauge transformation, we could alternatively reabsorb the Wilson line
phase $\alpha$ into the twist of the scalar field: $\Phi(2\pi R) = e^{2i\pi \alpha}\Phi(0)$.}
At one-loop order, the effective potential $V(\alpha,T)$ is given by one of the relations
reported in the appendix, eqs.(\ref{Pot2}), (\ref{Pot3}) or (\ref{Pot4}), with $d=3$, $M=0$, $\eta=0$ and a factor
of 2 taking into account the two scalar degrees of freedom.
At higher order, the leading contributions arise from the daisy diagrams drawn in figure \ref{figGaugeDaisy}.
Since $g_4\ll \lambda$, we can consistently neglect the daisy diagrams involving one-loop mass corrections
induced by the gauge field and consider only the one arising from the quartic coupling $\lambda$ (see figure 18).

\begin{figure}[h!]
\begin{center}
\includegraphics[width=.25\textwidth]{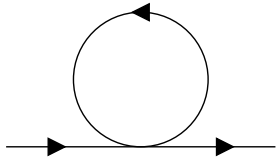}
\caption{The one-loop mass correction $M^2_{1-loop}$ to the charged scalar $\Phi$, induced by the
$\lambda (\Phi^\dagger \Phi)^2$ coupling.}\label{figThermalMass}
\end{center}
\end{figure}

Let us denote by $I_n$ the $(n+1)$-loop daisy diagrams obtained by summing all possible $n$ one-loop mass
insertions at zero external 5D momentum and define an effective parameter $\gamma(n)=I_{n}/I_{n-1}$;
if there exists some $n$ for which $\gamma(n) \sim 1$ or higher,
perturbation theory breaks down and all daisy diagrams must be resummed.
It is not difficult to compute $\gamma(n)$ for $n\gg 1$.
In this limit, one has to compute only the diagrams (b) in figure \ref{figGaugeDaisy},
since they have a combinatorial factor
proportional to $n$, whereas the diagrams (a) have a constant combinatorial factor.
Moreover, the $n$ one-loop mass insertions, namely the ``small'' bubbles in red in figure \ref{figGaugeDaisy},
decouple from the computation and one has only to compute the finite ``big'' bubble, the one in black
in figure \ref{figGaugeDaisy} (b).
We get
\be
\gamma(n) \simeq - \left[\frac{LM_{1-loop}(T,\alpha)}{2\pi}\right]^2
\frac{\sum_{m,k} \bigg[(k+\alpha)^2 + (LT)^2 m^2\bigg]^{\frac{-2n-1}2}}
{\sum_{m,k} \bigg[(k+\alpha)^2 + (LT)^2 m^2\bigg]^{\frac{-2n+1}2}}\,.
\label{gamma-def}
\ee
In eq.(\ref{gamma-def}), $M^2_{1-loop}(T,\alpha)$ is the one-loop finite thermal mass correction for the field $\Phi$ which,
for $\lambda\gg g$, is given only by the diagram in figure \ref{figThermalMass} and reads
\be
M^2_{1-loop}(T,\alpha) = \frac{(L T) T^2 \lambda}{4\pi^{2}} \sum_{\tilde k=1}^\infty \sum_{\tilde m=-\infty}^{\infty}
\frac{\cos(2\pi \alpha \tilde k)}{\bigg[\tilde m^2 + (LT)^2 \tilde k^2\bigg]^{3/2}}\,.
\label{M2-Phi}
\ee
The mass correction (\ref{M2-Phi}) is defined in a regularization in which only the finite terms arising from the compactification
are left, neglecting the UV divergence taken to be equal to that obtained in non-compact 5D space.
This is technically achieved by dropping the Poisson resummed $\tilde k=0$ term in eq.(\ref{M2-Phi}).
We immediately recognize from eq.(\ref{gamma-def}) that for $\alpha\simeq 0$, the leading contribution to $I_n$ is given
by the KK and Matsubara zero modes $k=m=0$, which give rise to $|\gamma(n)|\sim 1/\alpha^2$. The very low
values of the Wilson line phase $\alpha$ are the most affected by daisy diagrams, as is obvious by
the fact that IR divergences in the diagrams $I_n$ appear precisely in the limit $\alpha\rightarrow 0$.
A more transparent estimate of the relevance of resumming daisy diagrams can be performed
in the two extreme limits when $LT\ll 1$ and $LT \gg 1$. In these two limits, and for sufficiently high
$n$, $\gamma(n)\simeq \gamma$ is independent of $n$. For $LT\ll1$, we can approximate the double
sum in eq.(\ref{gamma-def})
by taking $k=0$ and substituting the sum over the Matsubara modes with an integral.
Viceversa, in eq.(\ref{M2-Phi}), we can take $\tilde m=0$, retain only the sum over the
modes $\tilde k$ and fix $\alpha=0$.\footnote{Notice that
for sufficiently high values of $\alpha$, $\alpha\gtrsim  0.2$, $M^2_{1-loop}(\alpha,T)$ becomes negative and
a non-vanishing VEV for the field $\Phi$ is induced, which complicates the above analysis. For this reason, we analyze
here only the region of low values of $\alpha$, which are anyway the ones most affected by the daisy diagrams.}
In this way, we get, for $LT\ll 1$,
\be
\gamma \simeq - \frac{\lambda \,\zeta(3)}{4 \pi^{2} L^2} \frac{1}{(\frac{\alpha}R)^2} = - \frac{M^2_{1-loop}(0,0)}{M^2_{tree}}\,,
\label{gamma-lowT}
\ee
where $M_{tree}^2=(\alpha/R)^2$ is the tree-level mass of the lowest KK mode for the field $\Phi$.
Daisy diagrams start to be relevant when the thermal mass is of the same order (or higher) than
the tree-level one.

For very high temperatures, $LT \gg 1$, the double sums in eq.(\ref{gamma-def}) are well approximated
by the leading term $k=m=0$, whereas in the mass correction (\ref{M2-Phi}) we can
approximate the sum over $\tilde m$ with an integral (setting again $\alpha=0$) and
sum over all $\tilde k$.
One gets, for $LT\gg 1$,
\be
\gamma \simeq - \frac{\lambda LT}{48 \pi^{2}} \frac{1}{\alpha^2}\,.
\label{gamma-highT}
\ee
As eq.(\ref{gamma-lowT}), eq.(\ref{gamma-highT}) can be written as the ratio of the thermal mass at high temperatures
over the tree-level mass $M^2_{tree}$. In fact, as a rough estimate, obtained by taking $k=m=0$
in the two double sums in eq.(\ref{gamma-def}), one sees that at any temperature
$\gamma \sim - M^2_{1-loop}/M^2_{tree}$.

Once established the necessity of considering higher-loop daisy diagrams for sufficiently low values of the
Wilson line phase $\alpha$, let us resum all these diagrams by following the
procedure outlined in  ref.~\cite{Parwani:1991gq}. Since $\gamma<0$, one can expect some cancellation between daisy diagrams
of different order, resulting in a decrease of the effect of the daisy resummation procedure.
This is indeed what happens, as we will see.
As briefly discussed before, along the lines of  ref.~\cite{Parwani:1991gq}, the resulting
one-loop improved effective potential $V_{imp}(T,\alpha)$  is obtained by
one of the relations reported in the appendix, eqs.(\ref{Pot2}), (\ref{Pot3}) or (\ref{Pot4}),
but by taking $M=M_{1-loop}(T,0)$ instead of $M=0$. From eq.(\ref{Pot4}), for instance, one gets
\be
V_{imp}(T,\alpha) = -\frac{4 L T^5}{(2\pi)^{5/2}} \sum_{\tilde k=1}^\infty
\sum_{\tilde m=-\infty}^\infty \frac{\cos(2\pi \tilde k \alpha)}{\big[\tilde m^2 + (LT\tilde k)^2\big]^{\frac 52}}
B_{5/2}\bigg(\frac{M_{1-loop}(T,0)}T \sqrt{\tilde m^2 + (LT \tilde k)^2}\bigg).
\ee

A good estimate of the higher-order daisy contributions at any temperature is performed by
comparing the Wilson line thermal mass obtained from $V_{imp}(T,\alpha)$ with that derived from
$V(T,\alpha)$. We denote them respectively $M^2_{imp}(T,\alpha)$ and $M^2(T,\alpha)$ (not to be confused
with $M^2_{1-loop}(T,\alpha)$ in eq.(\ref{M2-Phi}), which is the thermal mass of the field $\Phi$).
We plot in figure 19 (a) $M^2_{imp}(T,\alpha)$ and $M^2(T,\alpha)$ as function of $\alpha$
for $\lambda \sim 1$ and for $LT\sim 1$. As can be seen, the two behaviours are very similar,
with deviations which are at most of about 10$\%$ for $\alpha=0$. As expected,
the maximal deviation is found for values of $\alpha\simeq 0$.
We find that at low temperatures ($LT\sim 1/10$), the maximal deviations (at $\alpha=0$) are of order
of a few $\%$, whereas at high temperatures ($LT\sim 10$) they are of order $30\%$.
Notice that, although this model does not have a phase transition in the Wilson
line parameter, $LT\sim 1$ is in the range of the critical temperature where
the phase transition occurs in such kind of models, as we have explicitly seen
in sections 2 and 3. From our analysis we thus find that the leading higher-loop daisy diagrams
give small corrections to the effective potential. This can be taken as an evidence
of the fact that perturbation theory in our 5D models is still valid at $T\sim T_C$.
On the contrary, for very high temperatures $(LT >> 10)$,
the one-loop improved potential starts to significantly differ from the na\"\i ve one-loop
potential. This is particularly evident from the equivalent expression of the
potential given in eq.(\ref{Pot2}). At very high temperatures,
the na\"\i ve one-loop potential is entirely dominated by the Matsubara zero mode $m=0$, proportional to $T$,
whereas the one-loop improved potential, due to the non-vanishing argument in the function
$B_2$ appearing in eq.(\ref{Pot2}), is exponentially suppressed in $LM\simeq \sqrt{LT}$.
The discrepancy that one finds between the two potentials starts however to be very large
for temperatures which are well above the cut-off scale $\Lambda$.
As discussed in footnote \ref{footCutOff}, it is meaningless to consider such range of temperatures
in an effective field theory approach.

\begin{figure}[h!]
\begin{center}
\begin{tabular}{c c}
\includegraphics[width=.465\textwidth]{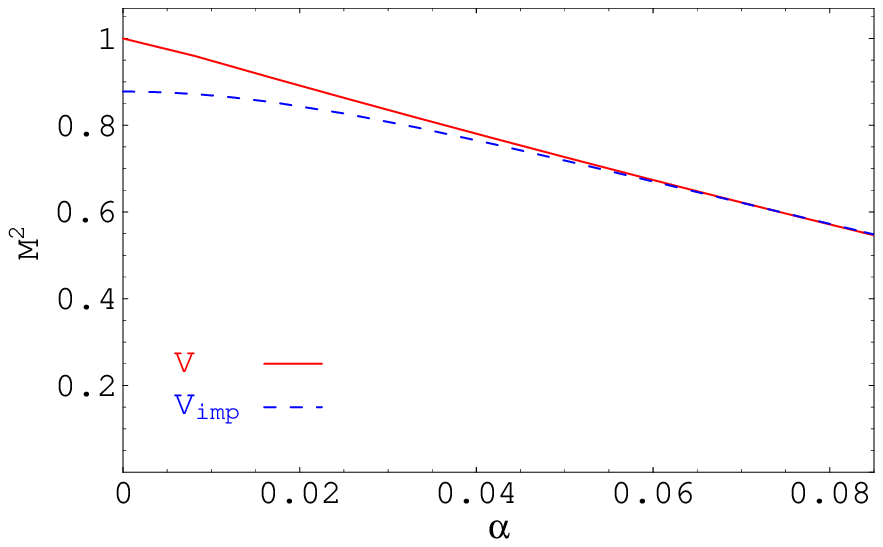}
&\includegraphics[width=.465\textwidth]{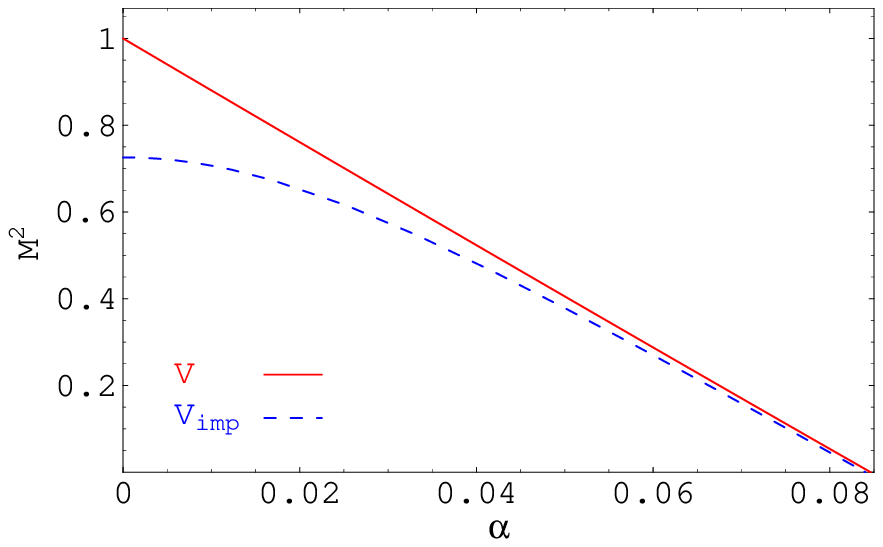}\\
(a) & (b)
\end{tabular}
\caption{(a) Thermal masses in 5D, from the one-loop na\"\i ve potential (red solid line)
and the one-loop improved one (blue dashed line). (b) Thermal masses in the reduced 4D model,
from the one-loop na\"\i ve potential (red solid line) and the one-loop improved one (blue dashed line).
In both cases the normalization is such that $M^2(\alpha=0,T)=1$.}\label{ThMass}
\end{center}
\end{figure}

It is useful to compare the contribution of daisy diagrams for the 5D $U(1)$ model above
with that of the 4D model arising by trivial dimensional reduction, where all the massive KK modes
are neglected.
By denoting with $a$ and $\phi$ the properly normalized zero mode components of $A_5$ and
$\Phi$, one gets the following 4D Lagrangian:
\be
{\cal L} = (D_\mu \phi)^\dagger D^\mu \phi +\frac{1}{2} (\partial_\mu a)(\partial^\mu a)
- g^2\, a^2\,\phi^\dagger \phi
-\frac{\lambda}{4}(\phi^\dagger \phi)^2 - \frac 14 F_{\mu\nu}F^{\mu\nu}\,.
\label{lag-Phi-a}
\ee
The trivial dimensional reduction spoils the 5D Wilson line nature of $\alpha$, breaking
the shift symmetry $\alpha\rightarrow \alpha+1$.
Let us study the effective potential $V(\alpha)$ for the VEV of the field $a$ ($\langle a \rangle =\alpha/(g R)$).
The contribution of $\Phi$ to $V(\alpha)$
is the standard one for a massive scalar field
at finite temperature \cite{Dolan:1973qd}. One has, neglecting terms independent of $\alpha$,
\be
V(T,\alpha) = V(0,\alpha)+\frac{T^4}{\pi^2}\int_0^\infty \! dx \,x^2
\log \bigg[1 - e^{-\sqrt{x^2 + (\alpha/RT)^2}}\bigg]\,.
\label{Pot-4d-a}
\ee
In eq.(\ref{Pot-4d-a}), $V(0,\alpha)$ is the zero temperature effective potential, which is UV divergent.
In the $\overline{{\rm MS}}$ regularizatrion scheme, it reads
\be
V(0,\alpha) = \frac{\alpha^4 \pi^2}{2 L^4} \bigg[\log\left(\frac{\alpha^2}{R^2 \mu^2}\right)-\frac 32\bigg]\,,
\ee
where $\mu$ is the renormalization scale.

The one-loop improved potential is obtained by substituting in eq.(\ref{Pot-4d-a}) the tree-level
mass with the one-loop improved one, computed in the 4D reduced model.
For $g\ll \lambda$, this simply amounts to shifting
$(\alpha/R)^2\rightarrow (\alpha/R)^2 + \lambda T^2/12$.
We plot the thermal masses obtained by the one-loop na\"\i ve and improved potentials in figure 19 (b) for $LT\sim 1$,
$\lambda\sim1$ and setting $\mu=1/R$, which is the scale above which our effective theory breaks down.
The comparison between figures 19 (a) and (b) clearly indicates that the
5D model is less sensitive to higher-loop effects than the 4D reduced theory, although
both models are not much sensitive to the daisy resummation procedure for $\alpha\gtrsim 0.05$.

Let us now comment on the daisy diagram contributions to
the Higgs effective potential in 5D gauge-Higgs unification models where no fundamental
scalars are present. Fermion loops do not give rise to IR divergencies since the Matsubara modes
are half integers and there is always an effective non-vanishing thermal mass. Moreover,
chiral symmetry implies that fermion mass corrections are always proportional to their tree-level
mass and thus they vanish in the IR limit of vanishing mass.
There is then no need of resumming daisy diagrams for fermions. Gauge invariance forbids the appearance
of any thermal mass correction to the transverse polarization of gauge bosons which are
also unaffected by the daisy resummation procedure. The Higgs field itself, $H$,
does not have tree-level self-interactions and hence does not give any higher-loop contribution
to its potential in the daisy approximation.
Finally, we are left with the longitudinal components $A_0$
of the gauge fields, that acquire a non-vanishing thermal mass correction (Debye mass);
the corresponding contributions to the Higgs potential have to be daisy-resummed.
We expect that the latter will be of the same order of magnitude as the one we have estimated
in our simple 5D model defined by the Lagrangian (\ref{lag-Phi}).

As it happens in the SM, we might expect that the resummation of daisy diagrams
in our 5D model leads to a change in the cubic term of the approximate
Wilson line potential (\ref{Pot-gen}). However, since transverse gauge bosons are not unaffected by daisy
diagrams, this shift would only lead to a slight change of the potential, which will still
predict a first-order phase transition.

As final remark, note that the absence of a tree-level potential in Wilson line based models permits to evade
Weinberg's rough argument for the necessity of the breakdown of perturbation theory
around the phase transition \cite{Weinberg:1974hy}: since the tree-level potential appearing
in generic quantum field theories is temperature independent, perturbative corrections
(which are instead temperature-dependent)
cannot give rise to a drastic change in the potential (as needed for a phase transition),
unless perturbation theory breaks down.\footnote{The possibility that perturbation theory
does not necessarily break down in our 5D model is also suggested by the value of the
critical temperature $T_C\sim 1/L$, which is independent on any coupling constant.
This has to be contrasted, for instance, with the SM case, in which roughly
$T_C\sim \sqrt{-\mu^2/\lambda}$, where $\mu^2$ is
the Higgs mass term.}

\section{Conclusions}

We have studied various aspects of the dynamics of Wilson line phases at
finite temperature. From a more theoretical viewpoint, we have first of all
shown that the one-loop effective potential is gauge invariant, contrary
to ordinary effective potentials in non-abelian gauge theories in 4D,
including the SM.
We have also pointed out that the non-local nature of the Wilson line
potential, which already ensures its finiteness
at zero temperature at all orders in perturbation theory, is responsible
for its mild dependence on quantum corrections at finite temperature.
We have established this result
by computing the one-loop improved thermal mass --- obtained by resumming
certain IR divergent higher loop diagrams --- with the na\"\i ve one in a given 5D model, and showing
that the improvement obtained by the resummation is small.
This gives strong evidence that in models in which the order parameter
is given by a Wilson line phase (models of gauge-Higgs unification),
perturbation theory generally holds at least up to the critical temperature $T_C\sim 1/L$.
Of course, it would be interesting and reassuring to establish
this result by performing a systematic study
of higher loop corrections for these 5D models.

At a more phenomenological level, we have studied how the
phase transition occurs in Wilson line based models, showing that this
is typically of first-order. We have then focused our attention to a particular
model \cite{Scrucca:2003ra}, in which the Wilson line phase transition is identified
with the electroweak phase transition.
In this model, unless some generalizations are considered, the Higgs mass is always predicted
to be too low: $M_{H}< M_W/2$.
In this range, the transition is strongly of first order,
similarly to what happens for the SM in the same regime of parameters.

Introducing bulk fermions in large representations of the $SU(3)_w$ gauge group allow
for reasonable Higgs masses. In this case, the first order phase transition is
very weak and most likely unable to give a successful electroweak baryogenesis.
Another possible option to increase the Higgs mass is obtained by
introducing large localized gauge kinetic terms.
Interestingly enough, the first order phase transition is now
considerably stronger than in the previous case and it might
even be sufficiently strong for electroweak baryogenesis.

We tacitly assumed in our paper that the dynamics of Wilson line phases
at finite temperature is decoupled from that of space-time itself, namely
that the mechanism stabilizing the internal direction does not alter
our analysis or takes place at temperatures higher than the critical temperature $T_C$.
Since the compactification scale is of the same order of $T_C$, it would be
interesting to see in some concrete model of radius stabilization
if this assumption can actually be verified.

\section*{Acknowledgments}

We would like to thank A. Hebecker, M. Quiros, L. Silvestrini and A. Wulzer
for useful discussions and comments.
This research work was partly supported by MIUR and by
the European Community through the RTN
Programs MRTN-CT-2004-005104 and MRTN-CT-2004-503369.

\appendix

\section{Explicit form of the potential}\label{appEffPot}

\subsection{Bulk fields contribution}

There are various ways of computing eq.(\ref{Pot0}) for bulk fields.
For generality, we will report in the following the result that one obtains
in $d$ non-compact dimensions.
A possible computation is to disentangle the $T=0$ and $T\neq 0$ terms in eq.(\ref{Pot0}) and compute
the $T$-dependent term as usually done in 4D for each KK state. In this case, one simply gets
(see {\it i.e.} \cite{Dolan:1973qd,Arnold:1992rz})
\be
V(T,q\alpha) = V(T=0,q\alpha) + (-)^{2\eta} \,T \sum_{k=-\infty}^{+\infty} \int\!\frac{d^dp}{(2\pi)^d}
\log \bigg[1 - (-)^{2\eta} e^{-\frac{\sqrt{p^2 + M_k^2}}T}\bigg] \,, \label{Pot1}
\ee
where $p^2$ denotes the $d$ dimensional non-compact momentum square.

Other useful and more explicit forms of the potential are obtained by using the relation ${\rm tr}\,\log A=
-\int_0^\infty dt/t \, {\rm tr}\,e^{-tA}$ in eq.(\ref{Pot0}), valid up to irrelevant constants.
Subsequently, one can use the Poisson resummation formula
\be
\sum_{n=-\infty}^{+\infty} e^{-\pi t (n+a)^2} = \frac{1}{\sqrt{t}} \sum_{\tilde n=-\infty}^{+\infty}
e^{-\frac{\pi \tilde n^2}t} e^{2i\pi \tilde n a}\,.
\label{poisson}
\ee
The above formula can be applied in eq.(\ref{Pot0}) to the KK modes $k$, to the Matsubara ones $m$,
or to both, resulting in different but equivalent ways of computing the effective
potential. In the three cases, one gets, respectively:
\be
V(T,q\alpha) = \frac{2T(-)^{2\eta-1}}{(2\pi)^{\frac{d+1}2}L^d}
\sum_{\tilde k=1}^\infty
\sum_{m=-\infty}^\infty \frac{\cos(2\pi \tilde k q\alpha)}{\tilde k^{d+1}}
B_{\frac{d+1}2}\bigg[L\tilde k\sqrt{M^2+[2\pi (m+\eta)T]^2}\bigg] \,, \label{Pot2}\\
\ee
where
\be
B_{\frac{d+1}2}(x) \equiv x^{\frac{d+1}2} K_{\frac{d+1}2}(x)
\ee
and $K_n$ are the modified Bessel functions.
By Poisson resumming over $m$, we get
\bea
V(T,q\alpha) &=& (-)^{2\eta-1}\frac{L}{\pi(2\pi)^{\frac d2}}\sum_{\tilde k=1}^\infty
\frac{\cos(2\pi \tilde k q\alpha)}{(L \tilde k)^{d+2}}
B_{\frac{d+2}2}(L M\tilde k) + \label{Pot3}\\
&& 2(-)^{2\eta-1}\sum_{k=-\infty}^\infty \sum_{\tilde m=1}^\infty (-)^{2\tilde m\eta}
\left(\frac{T}{\tilde m\sqrt{2\pi}}\right)^{d+1}\!\!
B_{\frac{d+1}2}\bigg(\frac{M_k \tilde m}T\bigg) \,, \nn
\eea
where a Poisson resummation over $k$ has been performed in the first term ($\tilde m=0$).
Finally, by Poisson resumming in both indices, one finds
\bea
V(T,q\alpha) &=& (-)^{2\eta-1}\frac{L T^{d+2}}{\pi(2\pi)^{\frac d2}}\sum_{\tilde k=1}^\infty
\sum_{\tilde m=-\infty}^\infty (-)^{2\eta \tilde m}
\bigg[\tilde m^2+(L T\tilde k)^2\bigg]^{-\frac{d+2}2} \nn \\
&& B_{\frac{d+2}2}\bigg[\frac MT\sqrt{\tilde m^2+(L T\tilde k)^2}\bigg]
\cos(2\pi \tilde k q\alpha) \,. \label{Pot4}
\eea
In eqs.(\ref{Pot2}),(\ref{Pot3}) and (\ref{Pot4}) an irrelevant $\alpha$-independent
term has always been omitted.
All these relations also hold for vanishing bulk mass term, $M=0$, by noticing that
$\lim_{x\rightarrow 0} B_\alpha(x) = 2^{\alpha-1}\Gamma(\alpha)$, where $\alpha$
is any positive integer or half-integer number.

\subsection{Boundary-bulk fields contribution}

The contribution to the Higgs effective potential of bulk and boundary fields mixed
as in eq.(\ref{Lagferm}) is substantially more involved than that of purely bulk fields.
One has essentially to compute $V(T,q\alpha)$ directly from
eq.(\ref{Pot0}), where $M_k^2$ are the mass eigenvalues of the bulk-boundary system.
It turns out that the full contribution can be written as a sum
of a purely bulk contribution and a remaining boundary term.
For the matter Lagrangian (\ref{Lagferm}), the bulk contribution is that given by
the bulk fermions $\Psi$ and $\tilde \Psi$ with $e_1=e_2=0$, whereas the boundary
contribution is a simple generalization of the one found in ref.~\cite{Scrucca:2003ra}
at $T=0$:
\bea
V_u(T,\alpha) &=& -\frac{8T}{\pi^2 L^3}\sum_{m=-\infty}^{+\infty}\int_0^\infty dx\, x^2
\ln\Bigg[\prod_{i=1}^{2} \mathrm{Re}\bigg[1+\delta_{i2}\frac{{\epsilon_2}^2}{2\widetilde{x}_\lambda}
f_0(\widetilde{x}_\lambda,0)
\nn\\ &&
+\frac{{\epsilon_i}^2}{2^{\delta_{i2}}\widetilde{x}_\lambda}
f_0(\widetilde{x}_\lambda,2\alpha)\bigg]
+\prod_{i=1}^2 \mathrm{Im}\bigg[\frac{{\epsilon_i}^2}{2^{\delta_{i2}}\widetilde{x}}
f_1(\widetilde{x}_\lambda,2\alpha)\bigg]\Bigg],\nn \\
V_d(T,\alpha) &=& -\frac{8T}{\pi^2 L^3}\sum_{m=-\infty}^{+\infty}\int_0^\infty dx\, x^2
\ln\Bigg[\prod_{i=1}^{2} \mathrm{Re}\bigg[1
+\delta_{i1}\frac{{\epsilon_1}^2}{2\widetilde{x}_\lambda} f_0(\widetilde{x}_\lambda,\alpha)\bigg]
\Bigg]\,.
\label{Vboundary}
\eea
In eq.(\ref{Vboundary}), $\lambda= M L/2$ and $\epsilon_{1,2}=\sqrt{L}e_i/2$ are dimensionless
parameters defined from the bulk mass $M$ and mixing terms $e_i$ appearing in eq.(\ref{Lagferm}),
\be\label{eqDefTempX}
\widetilde{x}^2 \equiv x_0^2 + x^2
\quad \textrm{and} \quad
\widetilde{x}_\lambda^2 \equiv (x_0^2 + x^2 + \lambda^2),
\ee
and $x_0 = (m + 1/2) \pi L T$ is the properly rescaled Matsubara frequence for a fermion.
The functions $f_{0,1}$ are defined as in ref.~\cite{Scrucca:2003ra}:
\begin{eqnarray}
f_0(\widetilde{x}_\lambda,\alpha)  & = &\sum_{n=-\infty}^\infty \frac 1{\widetilde{x}_\lambda + i \pi (n+\alpha)}
= \coth (\widetilde{x}_\lambda + i \pi \alpha) \;, \label{f0} \\
f_1(\widetilde{x}_\lambda,\alpha) & = & \sum_{n=-\infty}^\infty \frac {(-1)^n}{\widetilde{x}_\lambda +
  i \pi (n+\alpha)}
= \sinh^{-1} (\widetilde{x}_\lambda + i \pi \alpha) \;. \label{f1}
\end{eqnarray}

In presence of localized gauge kinetic terms, the gauge field contribution is given by the
sum of the purely bulk term and an additional boundary term. The latter, for a field
which couples diagonally with the Higgs field with charge $q$, can be written as
\be
V_g^{c_i}(T,q \alpha) = \frac{6 T}{\pi^2 L^3} \sum_{m=-\infty}^{+\infty}
\int_0^\infty\!\! dx\, x^2 \ln\left[
\prod_{i=1}^{2} \mathrm{Re}\left[1+c_i \widetilde{x} f_0(\widetilde{x},q \alpha)\right]
-\prod_{i=1}^{2} \mathrm{Re}\left[c_i \widetilde{x}f_1(\widetilde{x},q \alpha)\right]
\right],
\label{V-BKT}
\ee
with $\widetilde{x}$ defined as in eq.(\ref{eqDefTempX}), but with $x_0 = m \pi L T$.

\end{document}